\documentclass[fleqn,usenatbib]{mnras}

\usepackage[T1]{fontenc}

\DeclareRobustCommand{\VAN}[3]{#2}
\let\VANthebibliography\thebibliography
\def\thebibliography{\DeclareRobustCommand{\VAN}[3]{##3}\VANthebibliography}

\usepackage{graphicx}% Including figure files
\usepackage{amsmath,amssymb}	% Advanced maths commands
\usepackage{bm}
\usepackage{color}

\defcitealias{Linial&Metzger2023}{LM23}
\defcitealias{Strubbe&Quataert2009}{SQ09}

\title[QPEs in slim disk]{Quasi-Periodic Eruptions as a Probe of Accretion Disk in Tidal Disruption Events}

\author[T. Suzuguchi \& T. Matsumoto]{
Tomoya Suzuguchi$^{1}$\thanks{E-mail: suzuguchi@tap.scphys.kyoto-u.ac.jp (TS)} and 
Tatsuya Matsumoto$^{2,3,4}$
\\
$^{1}$Department of Physics, Kyoto University, Sakyo, Kyoto 606-8501, Japan\\
$^{2}$Department of Astronomy, Kyoto University, Sakyo, Kyoto 606-8501, Japan\\
$^{3}$The Hakubi Center for Advanced Research, Kyoto University, Kyoto 606-8501, Japan\\
$^{4}$Department of Astronomy, School of Science, the University of Tokyo, Bunkyo-ku, Tokyo 113-0033, Japan
}

\date{Accepted XXX. Received YYY; in original form ZZZ}

\pubyear{\the\year{}}

\begin{document}
\label{firstpage}
\pagerange{\pageref{firstpage}--\pageref{lastpage}}
\maketitle

% Abstract of the paper
\begin{abstract}
Quasi-periodic eruptions (QPEs) are X-ray transients characterized by nearly regular recurring flares from galactic nuclei. Recent observations have confirmed that some QPEs occur in galactic centers that experienced a tidal disruption event (TDE) a few years earlier. This may be reasonably explained if QPEs are produced when a star orbiting a supermassive black hole passes through an accretion disk formed by the TDE. Based on this scenario, we investigate the expected QPE signatures in the early stages of TDEs, taking into account the time evolution of the accretion disk. In the early phase, the disk is in a super-Eddington accretion state. The interaction between the star and such a slim disk results in QPEs with durations of $\sim 100-1000\,{\rm s}$ and temperatures of $\sim 1-100\,{\rm keV}$, which are significantly shorter and hotter than those of the currently detected QPE population. These events are detectable with current X-ray telescopes, but their small duty cycle ($\lesssim1\,\%$) and the potential presence of a massive disk wind may make detection challenging. We encourage early-time and long-term monitoring TDEs showing X-rays to capture these QPEs, as such detections would provide valuable insights into the disk formation process in TDEs.
\end{abstract}

% Select between one and six entries from the list of approved keywords.
% Don't make up new ones.
\begin{keywords}
accretion, accretion discs -- black hole physics -- transients: tidal disruption events -- X-rays: bursts
\end{keywords}

%%%%%%%%%%%%%%%%%%%%%%%%%%%%%%%%%%%%%%%%%%%%%%%%%%

%%%%%%%%%%%%%%%%% BODY OF PAPER %%%%%%%%%%%%%%%%%%

\section{Introduction}
Quasi-periodic eruptions (QPEs) are recurring X-ray transients in galactic nuclei with intervals ranging from a few hours to days. The first event was serendipitously discovered in 2018 during follow-up campaigns of an X-ray flare in the Seyfert 2 galaxy GSN 069 \citep{Miniutti+2019}, and a targeted observation of a galaxy similar to GSN 069 detected the second event \citep{Giustini+2020}. To date, about ten QPEs (including candidates) have been discovered through X-ray surveys \citep{Arcodia+2021,Arcodia+2024b,Arcodia+2025} and follow-ups of X-ray and optical transients \citep{Chakraborty+2021,Quintin+2023,Nicholl+2024,Chakraborty+2025a,Hernandez-Garcia+2025,Bykov+2025}. Each eruption of these objects has a duration of $10^{3}-10^{5}\,{\rm s}$ and the duty cycle is typically $\sim10\,\%$. The peak X-ray luminosity is  $10^{41}-10^{43}\,{\rm erg}\,{\rm s}^{-1}$ and the spectrum is fitted by a thermal one with temperature of $100-200\,{\rm eV}$. 

The origin of these periodic signals remains uncertain; however, they are likely associated with supermassive black holes (SMBHs) in galactic nuclei. Currently various theoretical models have been proposed, including an interaction between a star (including a compact object) and accretion disk around a SMBH \citep{Dai+2010, Xian+2021, Sukova+2021, Linial&Metzger2023, Franchini+2023, Tagawa&Haiman2023, Linial&Metzger2024b, Zhou+2024a, Zhou+2024b, Zhou+2025,Linial+2025,Tsz-Lok_Lam+2025,Vurm+2025,Yao+2025,Huang+2025}, mass transfer from a star to a SMBH  \citep{King2020,King2022,Krolik&Linial2022, Metzger+2022, Linial&Sari2023, Lu&Quataert2023, Olejak+2025, Yao&Quataert2025}, limit-cycle oscillations induced by disk instability \citep{Raj&Nixon2021, Pan+2022, Pan+2023, Pan+2025, Kaur+2023, Sniegowska+2023}, gravitational self-lensing by a SMBH \citep{Ingram+2021}, and Lense-Thirring precession of an outflow \citep{Middleton+2025}. 

Among the above interpretations, particular attention has been given to the interaction between a star and accretion disk. We hereafter refer to this as the ``EMRI+disk'' model, since the star may migrate toward the SMBH in an extreme mass-ratio inspiral (EMRI). In this scenario, X-ray flares are produced when the star (hereafter referred to simply as the EMRI) passes through the disk. One advantage of the EMRI+disk model is that the complex modulation of recurrence time can be explained by incorporating eccentricity of the EMRI and/or the general relativistic precession of the orbit \citep{Franchini+2023,Linial&Metzger2023}, which may be difficult to account for in the other models.\footnote{We note that detailed analyses of the emission process may suggest that a simple system composed of a star and an SMBH may have difficulty explaining the observations \citep{Guo&Shen2025,Linial+2025,Mummery2025,Tsz-Lok_Lam+2025}.} In addition, since EMRIs are expected to be promising targets of future space-based gravitational wave detectors, this model may become relevant in the context of multi-messenger astronomy \citep[e.g.,][]{Chen+2022, Kejriwal+2024, Lyu+2024, Duque+2025, Olejak+2025, Suzuguchi+2025, Lui+2025}. 

Recently, there is growing evidence that some QPEs are connected to tidal disruption events (TDEs), where a star is disrupted by the strong tidal forces of a SMBH during a close encounter \citep{Hills1975,Rees1988}. Observationally, TDEs are detected as luminous flares across various wavelengths in galactic centers \citep[e.g.,][]{Gezari2021}. Several QPEs have been identified in X-ray follow-up observations of flares considered likely to be TDEs \citep{Miniutti+2019,Chakraborty+2021,Quintin+2023,Nicholl+2024,Chakraborty+2025a,Bykov+2025}. In addition, host galaxies of QPEs and TDEs exhibit common characteristics, such as being preferentially found in post-starburst galaxies \citep{Wevers+2022,Wevers+2024}. The most compelling evidence for their connection is the discovery of QPEs in follow-up observations of the optical TDE AT2019qiz \citep{Nicholl+2024}.

In the context of the EMRI+disk scenario, a TDE resulting from the disruption of another star supplies gas that forms a disk, enabling interaction with the EMRI \citep{Franchini+2023,Linial&Metzger2023,Tagawa&Haiman2023}. Since the QPEs have typically been discovered several years after the associated TDEs, the accretion disk at that stage is reasonably described by a radiatively efficient, geometrically thin, so-called standard accretion disk \citep{Shakura&Sunyaev1973,Cannizzo+1990}. Thus far, most studies have focused on how the interaction between the EMRI and the standard disk could reproduce the observed properties of QPEs \citep[e.g.,][]{Chakraborty+2025b,Vurm+2025,Tsz-Lok_Lam+2025,Yao+2025,Guolo+2025}.

In this paper, based on the EMRI+disk scenario, we propose that QPEs can serve as probes of TDEs. Although TDEs have been extensively studied, some key aspects remain uncertain, especially how and when an accretion disk forms and contributes to the observed emission. Since the formation of the disk is a crucial process, it is important to investigate the expected observational signatures of QPEs that may emerge during this phase, and to discuss at what stage after the onset of a TDE such QPEs might appear. Although the details of disk formation are still uncertain, the accretion rate is expected to be high immediately after formation, making it necessary to consider not a standard thin disk, but rather a so-called slim disk \citep[e.g.,][]{Abramowicz+1988,Beloborodov1998,Wang&Zhou1999,Watarai&Fukue1999,Watarai2006}.

We organize this paper as follows. In Section~\ref{sec:emission}, we review the EMRI+disk model of QPEs. Section~\ref{sec:disk} introduces two slim disk models associated with TDEs, and Section~\ref{sec:evolution} discusses the time evolution of QPE properties based on each model. We discuss the parameter dependence of the properties and the detectability of early-time QPEs in Section~\ref{sec:discussion}, and summarize our findings in Section~\ref{sec:conclusions}.

\section{Emission model} \label{sec:emission}
We review the QPE emission process in the EMRI+disk model largely based on \cite{Linial&Metzger2023} (hereafter \citetalias{Linial&Metzger2023}) and estimate the QPE duration, luminosity, and temperature. In this model, QPEs arise from collisions between a star orbiting a central SMBH (EMRI) and an accretion disk surrounding the SMBH. When the EMRI crosses the disk supersonically, it produces shocks and compresses the gaseous medium, which subsequently expands above and below the disk. Photons generated inside the ejected material are initially trapped but eventually diffuse out to produce observed QPEs. 

\begin{figure}
    \centering
    \includegraphics[keepaspectratio, scale=0.2,bb=0 0 1087 887]{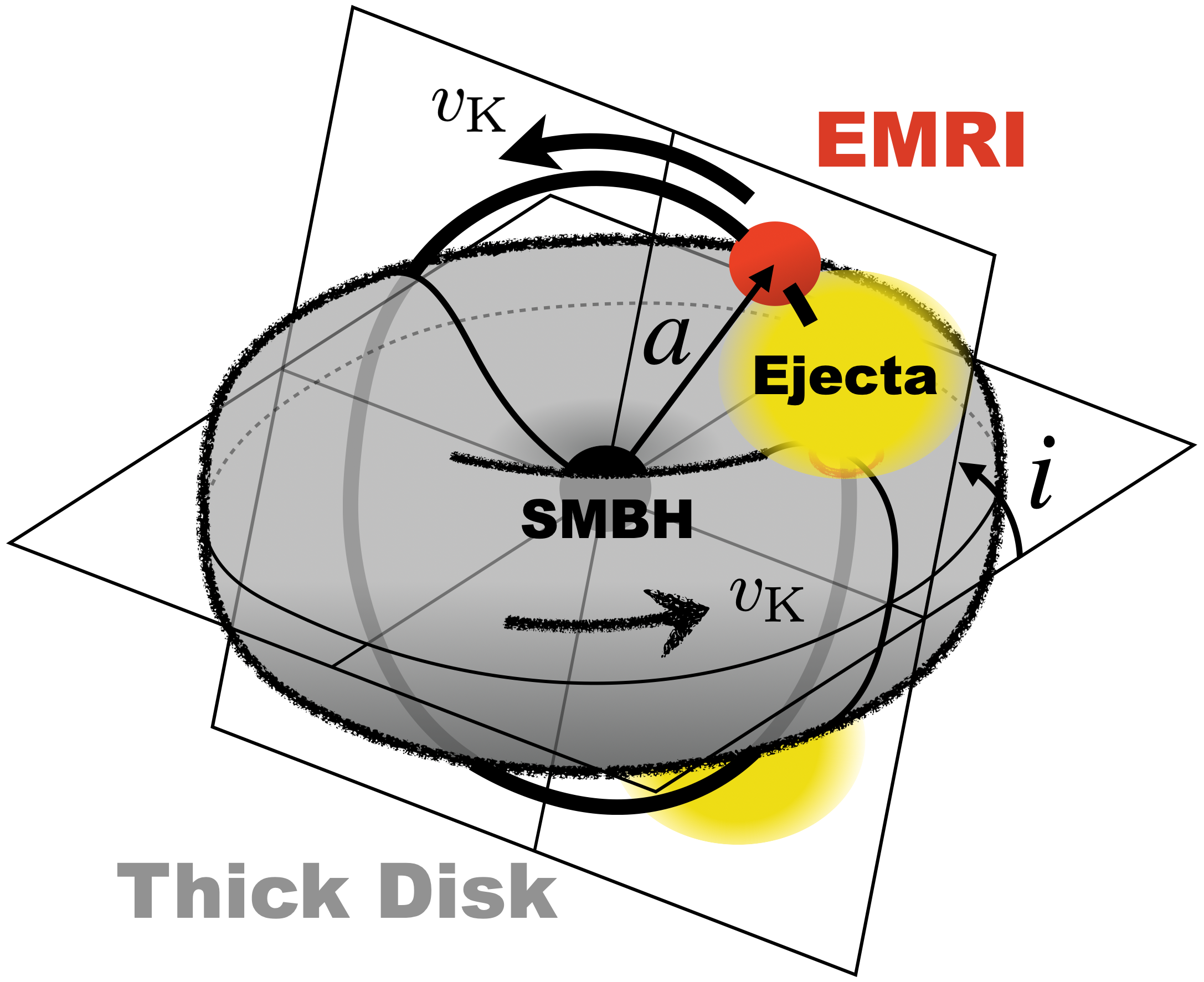}
    \caption{Schematic picture of an EMRI and thick disk system. The EMRI orbits around a SMBH with a semimajor axis $a$. The orbital plane is inclined from the equatorial plane by an angle $i$.}
    \label{fig:schematic}
\end{figure}

Figure~\ref{fig:schematic} shows a schematic picture of the EMRI+disk system. The main difference of our model from previous works is that instead of standard disks we consider super-Eddington accretion disks, which likely represent an early stage of the disk in TDEs \citep{Strubbe&Quataert2009}. The disk cannot cool efficiently and becomes geometrically thick as shown in Fig.~\ref{fig:schematic}, the so-called  ``slim disk'' \citep[e.g.,][]{Abramowicz+1988}. The distinct nature of the slim disk from the standard disk leads to a remarkable difference in the QPE signal. 

The EMRI with radius $R_{\star} = R_{\odot}{\cal R}_{\star}$ orbits the SMBH with mass $M_{\bullet} = 10^6M_{\odot} M_{\bullet,6}$, where $R_{\odot}$ and $M_{\odot}$ are the solar radius and mass, respectively. The orbital plane is inclined from the disk's midplane with an angle of $i$ ($0 \leq i \leq \pi$). We define that the EMRI is in prograde rotation when $i=0$ (see Fig.~\ref{fig:schematic}). The semimajor axis of the EMRI is $a$. Although the EMRI probably has a small eccentricity of $\sim 0.1$ to explain the observed long-short recurrence time, we consider a circular orbit for simplicity. 

\begin{figure}
    \centering
    \includegraphics[keepaspectratio, scale=0.35,bb=0 0 648 648]{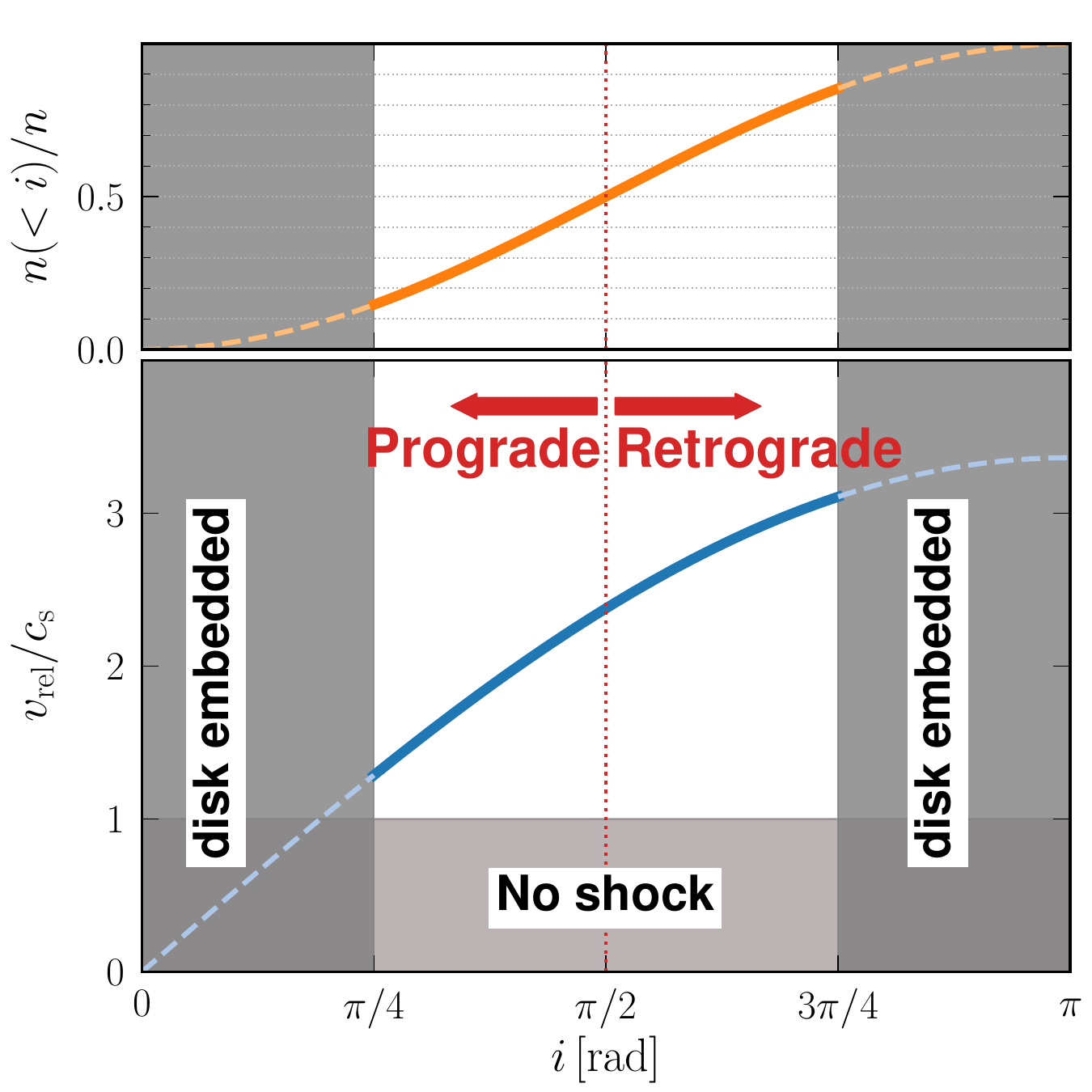}
    \caption{{\bf (Top)} Normalized cumulative number distribution of EMRIs as a function of the inclination angle $i$. EMRIs are assumed to distribute isotropically around the disk axis. EMRIs are in a prograde (retrograde) orbit for $i < (>) \pi/2$, and embedded in a thick disk for $i < \pi/4$ and $> 3\pi/4$ (shaded regions). {\bf (Bottom)} The ratio of relative velocity between a EMRI and disk material to the disk sound speed for different inclination. A shock is not formed when the ratio is smaller than unity (the horizontal shaded region).}
    \label{fig:inc_vel_num}
\end{figure}

Before calculating the detailed emission properties, we consider the condition to produce QPEs in the thick disk. Clearly, the EMRI should emerge from the disk during its orbital motion. This requires that the inclination angle should satisfy $\pi/4 \leq i \leq3\pi/4$; otherwise the EMRI is embedded within the disk. Here we assume that the disk's aspect ratio is about unity. Under this condition, we can estimate the fraction of EMRIs producing QPEs by assuming the direction of EMRI's angular momentum is distributed uniformly, $dn/d\Omega = 1/4\pi$. The cumulative number distribution of EMRIs is simply given by
\begin{align}
    \frac{n(<i)}{n} &= \int_{0}^{2\pi} {\rm d}\varphi \int_{0}^{i} \frac{{\rm d}n}{{\rm d}\Omega} \sin i^\prime {\rm d}i^\prime=\frac{1-\cos i}{2}\ ,
\end{align}
where $\varphi$ is the azimuthal angle around the vertical axis of the disk. The upper panel of Fig.~\ref{fig:inc_vel_num} shows the distribution. We find a good fraction of EMRIs ($\simeq 70\,\%$), satisfies the inclination condition. 

In what follows, we estimate the QPE properties based on \citetalias{Linial&Metzger2023}, whose model is applicable to other disk models than the standard thin disk. Here we retain explicit dependencies of QPE observables on disk properties so that we can apply the formulae to several thick disk models in Sec.~\ref{sec:disk}. The relative velocity between the EMRI and disk during the disk-crossing is given by $v_{\rm rel} = 2\sin{(i/2)}v_{\rm K}$ \citep[e.g.,][]{Murray&Dermott1999,Generozov&Perets2023}, where $v_{\rm K}=(GM_{\bullet}/a)^{1/2}$ is the Keplerian velocity and $G$ is the gravitational constant. Since the EMRI crosses the disk twice per orbital period, the QPE recurrence time is given by
\begin{align}
    P_{\rm QPE} &= \frac{P_{\rm orb}}{2} \simeq 4.3\,{\rm hr} \,
    \frac{a_{2}^{3/2}}{M_{\bullet,6}^{1/2}}\ ,
    \label{eq:P_QPE}
\end{align}
where $P_{\rm orb} = 2\pi a/v_{\mathrm{K}}$ is the orbital period of the EMRI and we normalize $a = 10^2R_{\rm g}a_{2}$. Here $R_{\rm g}=GM_{\bullet}/c^2$ is the gravitational radius, and $c$ is the speed of light. At each disk-crossing, the EMRI interacts with the disk material, intercepting gas with mass 
\begin{align}
    M_{\rm ej} &\simeq 2\pi\left(\frac{R_{\star}^{2}}{\sin{i}}\right) \Sigma \simeq 1.5 \times 10^{-8} M_{\odot}
    \frac{{\cal R}_{\star}^{2} \Sigma_{3}}{\sin{i}}\ ,
\end{align}
where $\Sigma = 10^{3}\,{\rm g} \, {\rm cm}^{-2} \Sigma_{3}$ is the disk surface density and its normalization is motivated by the thick disk (see below). The prefactor of $2$ accounts for inflows from the rotating disk \citepalias[see][]{Linial&Metzger2023}. 

The collision between the EMRI and disk results in a shock formation and ejection of materials. However, in contrast to the thin disk case, where the EMRI's orbital velocity is sufficiently higher than the sound speed of the pre-shock disk material, the relative orbital velocity can be comparable to the sound speed for the hot slim disk. To see this more quantitatively, we estimate the sound velocity of the slim disk from the vertical hydrostatic equilibrium, 
\begin{align}
    c_{\rm s} \simeq \left(\frac{P}{\rho}\right)^{1/2}
    \simeq \left[\frac{GM_{\bullet} H^2}{(R^2 + H^2)^{3/2}}\right]^{1/2} 
    \overset{H\simeq R} {\simeq} \frac{v_{\rm K}}{2^{3/4}}\ ,
\end{align}
where $P$ and $\rho$ are the disk's midplane pressure and density, respectively, and $H$ is the disk's scale height. The bottom panel of Fig.~\ref{fig:inc_vel_num} shows the ratio of the relative to sound velocities, that is the Mach number, for different inclination angles. The collision becomes supersonic for inclinations larger than $i \gtrsim 2\sin^{-1}(2^{-7/4}) \simeq 0.6$. This angle is smaller than $\pi/4$, and hence the non-embedded EMRIs automatically have a (marginally) supersonic orbital velocity and generate shocks by the collision with the disk. The shocked material expands below and above the disk with a velocity comparable to the shock and relative velocities, $v_{\rm ej} \simeq v_{\rm rel}$ \citep[e.g.,][]{Ivanov+1998}.

The shock-dissipated energy is equally partitioned into the internal and kinetic energies of the ejecta. The initial internal energy is then given by
\begin{align}
    \label{eq:ejecta_energy}
    E_{\rm ej} &\simeq \frac{1}{2}M_{\rm ej} v_{\rm ej}^2 
    \simeq 1.4 \times 10^{44} \, {\rm erg} \,
    \frac{\sin^{2}{(i/2)}}{\sin{i}}\frac{{\cal R}_{\star}^{2} \Sigma_{3}}{a_{2}}\ .
\end{align}
In the first few dynamical timescales, the ejecta undergoes acceleration due to $P {\rm d}V$ work; however, its terminal velocity is at most roughly $\sqrt{2}$ times larger than the initial one. Therefore, we ignore the acceleration and assume that the ejecta expands at the constant (initial) velocity. 

Initially the shocked material may have a cylindrical shape and its volume is estimated by (see \citetalias{Linial&Metzger2023})
\begin{align}
    {\cal V}_{\rm sh} \simeq \pi {R_{\star}}^2\frac{H}{7\sin i}\ .
\end{align}
The shocked gas expands isotropically, and at sufficiently late times its shape approaches a spherical one with radius of $\sim v_{\rm ej}t$. However, what we find below is that the photon diffusion occurs at an early stage where the initial size cannot be neglected. To take this into account, we approximate the ejecta volume as a cylinder with a radius $R(t)=v_{\rm ej}t+R_\star$ and height $H(t) = v_{\rm ej}t + H/7\sin{i}$, as ${\cal V}_{\rm ej} = \pi R(t)^{2} H(t)$. With this definition, the ejecta volume satisfies ${\cal V}_{\rm ej}(t=0)={\cal V}_{\rm sh}$. The ejecta density is given by $\rho_{\rm ej} = M_{\rm ej}/{\cal V}_{\rm ej}$.

The evolution of the shocked material is similar to the supernova ejecta \citep[e.g.,][]{Arnett1980}. Initially photons remain trapped in the opaque ejecta until the diffusion timescale becomes comparable to the dynamical timescale. This occurs at a timescale of 
\begin{align}
    \label{eq:duration}
    t_{\rm QPE} 
    &\simeq \left(\frac{\kappa_{\rm T} M_{\rm ej}}{4\pi c v_{\rm ej}}\right)^{1/2}
    \simeq 2.7 \times 10^{-2} \, {\rm hr} \,
    \frac{{\cal R}_{\star} \Sigma_{3}^{1/2} a_{2}^{1/4}}{(\sin{i} \sin{(i/2)})^{1/2}}\ ,
\end{align}
where $\kappa_{\rm T} = 0.34 \, {\rm cm}^{2} \, {\rm g}^{-1}$ is the Thomson opacity. This timescale corresponds to the QPE duration. Importantly this timescale can be shorter than the dynamical timescale of the ejecta, or expansion timescale along the direction of the EMRI's motion:
\begin{align}
    t_{\rm exp} \simeq \frac{H}{7v_{\rm ej}\sin{i}}
    \simeq 0.20 \, {\rm hr} \, \frac{H_{2} M_{\bullet,6} a_2^{1/2}}{\sin{i} \sin{(i/2)}}\ .
\end{align}
which potentially invalidates Eq.~\eqref{eq:duration}, which is obtained by assuming the negligible dynamical timescale. However, we find that even in this case, the emission timescale is given by Eq.~\eqref{eq:duration}, and the emission happens when the EMRI arrives at the disk surface like the shock breakout emission in supernovae.

The radiated energy is estimated by the internal energy at the diffusion time taking the adiabatic loss into account:
\begin{align}
    E_{\rm QPE} &\simeq 
    \left(\frac{{\cal V}_{\rm sh}}{{\cal V}_{\rm ej}(t_{\rm QPE})}\right)^{1/3} E_{\rm ej} 
    \nonumber \\
    &\simeq 1.0 \times 10^{44} \, {\rm erg} \,
    \frac{\sin^{3/2}{(i/2)}}{\sin^{5/6}{i}}
    \frac{{\cal R}_{\star}^{5/3} \Sigma_{3}^{1/2} H_{2}^{1/3} M_{\bullet,6}^{1/3}}{a_{2}^{3/4}}\ ,
\end{align}
We assume that the ejecta is radiation-pressure dominated. Note that, in the second line, we also assume that the diffusion timescale is longer than the dynamical timescale; otherwise, $E_{\rm QPE} \simeq E_{\rm ej}$. The QPE luminosity is given by the above radiated energy divided by the duration,
\begin{align}
    \label{eq:luminosity}
    L_{\rm QPE} &\simeq 1.1 \times 10^{42} \, {\rm erg} \, {\rm s}^{-1} \, 
    \frac{\sin^{2}{(i/2)}}{\sin^{1/3}{i}}
    \frac{{\cal R}_{\star}^{2/3} H_{2}^{1/3} M_{\bullet,6}^{1/3}}{a_{2}}\ .
\end{align}

The characteristic radiation temperature is estimated through two different approaches, depending on whether the radiation is in the thermal equilibrium with ejecta material or not. If the photon production and absorption are efficient, the thermal equilibrium is achieved, and the temperature is given by the blackbody one
\begin{align}
    T_{\rm BB} &= \left(\frac{u_{\gamma}}{a_{\rm rad}}\right)^{1/4} 
    \simeq 40 \, {\rm eV} \,
    \sin^{2/3}{i}
    \frac{H_{2}^{1/12} M_{\bullet,6}^{1/12}}{{\cal R}_{\star}^{1/3} \Sigma_{3}^{1/4}}\ ,
\end{align}
where $u_{\gamma} = L_{\rm QPE}/[4 \pi (v_{\rm ej} t_{\rm QPE})^{2} v_{\rm ej}]$ is the radiation energy density at $t \simeq t_{\rm QPE}$, and $a_{\rm rad}$ is the radiation constant. However, such a strong coupling is not always realized, and in fact the radiation is out of thermal equilibrium. To see this, we consider the so-called ``thermalization efficiency'' following \cite{Nakar&Sari2010}:
\begin{align}
    \label{eq:eta_def}
    \eta &= \frac{n_{\rm BB}(T_{\rm BB,sh})}{{\rm min}(t_{\rm exp},t_{\rm QPE})\dot{n}_{\gamma,{\rm ff}}(T_{\rm BB,sh},\rho_{\mathrm{sh}})}\ ,
\end{align}
which is the ratio of the photon number density required for thermal equilibrium, 
$n_{\rm BB}(T_{\rm BB})=a_{\rm rad} T_{\rm BB}^{4}/(3k_{\rm B}T_{\rm BB})$, to the photon number density produced by the free-free emission during the initial expansion timescale. Here $k_{\rm B}$ is the Boltzmann constant, and $T_{\rm BB,sh} \simeq (3\rho_{\rm sh} v_{\rm K}^{2}/a_{\rm rad})^{1/4}$ and $\rho_{\rm sh} \simeq 7\rho$ are the blackbody temperature and density of the ejecta immediately after the shock passage,\footnote{The photon production is most efficient at the beginning of the ejecta expansion because the density is highest at this moment (\citetalias{Linial&Metzger2023}).} with the pre-shock gas density of $\rho \simeq \Sigma/(2H)$. Using the free-free photon production rate of $\dot{n}_{\gamma}(T,\rho) \simeq 3.5 \times 10^{36} \, {\rm s}^{-1} \, {\rm cm}^{-3} \, \rho^{2} T^{-1/2}$ \citep[see][]{Nakar&Sari2010}, we obtain  
\begin{equation}
\label{eq:eta}
    \begin{split}
        \eta & \simeq 
        \begin{cases}
            \displaystyle
            21 \,
            \sin{i}
            \frac{H_{2}^{1/8} M_{\bullet,6}^{1/8}}{\Sigma_{3}^{9/8} a_{2}^{11/8}}
            &:\, t_{\rm QPE} > t_{\rm exp}\ , \\
            \\
            \displaystyle
            150 \,
            (\sin{i}\sin{(i/2)})^{1/2}
            \frac{H_{2}^{9/8} M_{\bullet,6}^{9/8}}{{\cal R}_{\star} \Sigma_{3}^{13/8} a_{2}^{9/8}}
            &:\, t_{\rm QPE} < t_{\rm exp}\ . \\
        \end{cases}
    \end{split}
\end{equation}
The parameter larger than unity $\eta > 1$ suggests that photon production is indeed inefficient, leading to a characteristic photon temperature higher than the blackbody one.

In addition, Comptonization also plays a role to determine the temperature, which was found to be minor in the context of observed QPEs (\citetalias{Linial&Metzger2023}, but see \citealt{Mummery2025}). Comptonization upscatteres low energy photons to $\sim k_{\rm B}T$ and increases the photon number available to thermalization \citep{Nakar&Sari2010}. The minimum energy of photons contributing to thermalization is determined by two processes \citep{Faran&Sari2019,Irwin&Hotokezaka2024}: One is the expansion of ejecta. Neglecting free-free absorption, the total number of scatterings within the single dynamical timescale is $\sim \kappa_{\rm T} \rho_{\rm sh} c [{\rm min}(t_{\rm exp},t_{\rm QPE})]$, and hence the Compton y-parameter is ${\cal Y} = 4(k_{\rm B}T/m_{\rm e}c^{2}) \kappa_{\rm T} \rho_{\rm sh} c [{\rm min}(t_{\rm exp},t_{\rm QPE})]$. The minimum energy is obtained by requiring $h \nu_{\rm exp} {\rm e}^{{\cal Y}} \gtrsim k_{\rm B} T$ (photons with $\nu < \nu_{\rm exp}$ cannot be upscattered to $\sim k_{\rm B} T$), where $h$ is the Planck constant. The other process is free-free absorption. Photons with an absorption optical depth $\tau_{\rm abs} = \kappa_{\rm ff,\nu} \rho_{\rm sh} c [{\rm min}(t_{\rm exp},t_{\rm QPE})] \gtrsim 1$ are absorbed before participating to thermalization. Here $\kappa_{\rm ff,\nu}$ is the free-free absorption opacity \citep[e.g.,][]{Rybicki&Lightman1986}. We denote the corresponding minimum energy as $\nu_{\rm abs}$. The minimum energy is given by the larger one from the above energies, $\nu_{\rm min} = {\rm max}(\nu_{\rm exp},\,\nu_{\rm abs})$. A factor accounting for the additional photons provided by Comptonization is then given by \cite{Nakar&Sari2010}
\begin{align}
    \label{eq:xi}
    \xi = {\rm max} \left[1,\frac{1}{2}\ln{y_{\rm max}}(1.6+\ln{y_{\rm max}})\right]\ ,
\end{align}
where $y_{\rm max} \equiv k_{\rm B}T/h\nu_{\rm min}$. 

With the effects of the photon generation and Comptonization, the characteristic temperature at the shock passage $T_{\rm sh}$ is ultimately given by 
\begin{align}
    \label{eq:temperature_comptonization}
    T_{\rm sh} \xi(T_{\rm sh})^{2} = \eta(T_{\rm BB,sh})^{2} T_{\rm BB,sh}\ .
\end{align}
Note that this is an implicit equation for the temperature. Since the thermalization is most efficient just after the shock passage, we can use $\eta$ and $\xi$ at this moment and obtain the characteristic temperature at the diffusion time
\begin{equation}
\label{eq:obs_temperature}
    \begin{split}
    \left(\frac{\eta}{\xi}\right)^2T_{\rm BB}
    &\simeq
    \begin{cases}
        \displaystyle
        0.68 \, {\rm keV} \,
        \sin^{8/3}{i} \\
        \displaystyle
        \quad \times 
        \frac{H_{2}^{1/3} M_{\bullet,6}^{1/3}}
        {{\cal R}_{\star}^{1/3} \Sigma_{3}^{5/2} a_{2}^{11/4}}
        \left(\frac{\xi}{5}\right)^{-2} 
        &:\, t_{\rm QPE} > t_{\rm exp}\ , \\
        \displaystyle
        37 \, {\rm keV} \,
        \sin^{5/3}{i}\sin{(i/2)} \\
        \displaystyle
        \quad \times 
        \frac{H_{2}^{7/3} M_{\bullet,6}^{7/3}}
        {{\cal R}_{\star}^{7/3} \Sigma_{3}^{7/2} a_{2}^{9/4}}
        \left(\frac{\xi}{5}\right)^{-2}
        &:\, t_{\rm QPE} < t_{\rm exp}\ , \\
    \end{cases}
    \end{split}
\end{equation}
which predicts hard X-ray QPEs at the early stage of the EMRI and disk interaction. Here we use $\xi=5$ for a typical value obtained by our fiducial parameters. In summary, the observed QPE temperature is given by
\begin{align}
    \label{eq:obs_temperature2}
    T_{\rm QPE}={\rm max}\left[1,\left(\frac{\eta}{\xi}\right)^2\right]T_{\rm BB}\ .
\end{align}
As pointed out in \citet{Nakar&Sari2010}, this prescription for estimating the temperature is valid only to $T_{\rm QPE} \lesssim 50 \, {\rm keV}$. At higher temperatures, electron-positron pair production becomes significant and suppresses further temperature increase \citep[e.g.,][]{Katz+2010,Budnik+2010}. Accordingly, we adopt $50 \, {\rm keV}$ as a conservative upper limit of $T_{\rm QPE}$.

\section{Accretion disk formed by TDE} \label{sec:disk}
We consider evolution of an accretion disk formed after a TDE, which interacts with an EMRI to produce QPEs. That being said, the disk formation in TDEs is not yet fully understood and remains a hot topic in TDE research mainly because of complicated hydrodynamics of bound debris \citep[see e.g.,][]{Bonnerot&Stone2021}. A star is disrupted by a SMBH when it approaches the BH closer than the tidal radius,
\begin{align}
    \label{eq:tidal_radius}
    R_{\rm T} &\equiv 
    \left(\frac{M_{\bullet}}{M_{\star,{\rm TDE}}}\right)^{1/3} 
    R_{\star,{\rm TDE}} 
    \simeq 47R_{\rm g}
    \frac{{\cal R}_{\star,{\rm TDE}}}
    {M_{\bullet,6}^{2/3} {\cal M}_{\star,{\rm TDE}}^{1/3}}\ ,
\end{align}
where $M_{\star,{\rm TDE}} = M_{\odot} {\cal M}_{\star,{\rm TDE}}$ and $R_{\star,{\rm TDE}} = R_{\odot}{\cal R}_{\star,{\rm TDE}}$ are the mass and radius of the disrupted star, respectively. After the disruption, half of the stellar debris is bound and falls back to the SMBH at a rate of \citep{Rees1988,Phinney1989}
\begin{align}
    \label{eq:fallback_rate}
    \dot{M}_{\rm fb} \simeq \frac{M_{\star,{\rm TDE}}}{3t_{\rm fb}}
    \left(\frac{t}{t_{\rm fb}}\right)^{-5/3},
\end{align}
where the fallback time is given by
\begin{align}
    \label{eq:fallback_time}
    t_{\rm fb} \simeq 
    41 \, {\rm day} \, 
    M_{\bullet,6}^{1/2} {\cal R}_{\star,{\rm TDE}}^{3/2}
    {\cal M}_{\star,{\rm TDE}}^{-1}\ .
\end{align}

Theoretically, a disk is expected to form when the bound debris comes back to the BH and undergoes dissipative processes. In the classical picture of \cite{Rees1988}, the relativistic orbital precession causes such dissipation and results in a prompt disk formation. Indeed numerical simulations confirmed this for the case that the stellar pericenter is sufficiently small, $R_{\rm p} \sim R_{\rm g}$ \citep{Hayasaki+2013,Bonnerot+2016,Hayasaki+2016}. However, a majority of TDEs likely happen with the pericenter comparable to the tidal radius, $R_{\rm p} \sim R_{\rm T}$. In this case, the falling back debris does not experience an efficient dissipation and goes away to collide with other incoming fresh debris at an apocenter of $\sim 10^{15} \, {\rm cm} \sim 10^{4} \, R_{\rm g} \, M_{\bullet,6}^{-1}$ \citep{Shiokawa+2015,Ryu+2023b,Steinberg&Stone2024,Price+2024}. The subsequent debris's evolution is still not clear and may form a large-scale envelope covering the system \citep[e.g.,][]{Metzger2022,Krolik+2025}, which likely settles into an accretion disk later. 

Observationally, the disk formation is probably probed by X-ray emissions. TDEs discovered via X-ray observations constitute a good fraction of TDEs observed to date \citep{Komossa2015,Sazonov+2021,Grotova+2025}, although their detailed evolution is not clear because of relatively sparse data. Roughly $\simeq 40\,\%$ of optically-discovered TDEs also show X-ray emissions \citep[e.g.,][]{Guolo+2024}. Their X-ray light curves are diverse at around the optical peak, but become stable at later phase indicating a disk formation. In addition, at sufficiently late time of $\sim 1000 \,{\rm days}$, UV and optical light curves also level off \citep{van_Velzen+2019,Mummery+2024}. Such ``plateaus'' are likely produced when the emission is dominated by slowly evolving accretion disk.

To summarize, the late-time observations suggest that the disk is present, but its early time evolution is still unclear, which prohibits us from modeling a precise early-time evolution of QPEs. However, regardless of the detailed formation process, the new-born TDE disk likely accretes materials at a super Eddington rate \citep{Strubbe&Quataert2009,Shen&Matzner2014}:
\begin{align}
    \dot{M}_{\rm Edd}=\frac{4\pi GM_\bullet}{\epsilon \kappa_{\rm T} c}\ ,
\end{align}
where $\epsilon=0.1$ is the radiative efficiency. Such a disk has a different properties from the standard thin disks which have been considered in the context of QPEs. Hence as the first step, we simply replace the thin disk with the super-Eddington disk and study how the QPE properties are altered. In the following we take two representative models of super-Eddington accretion disks, which hopefully capture the nature of disks in TDEs.

\subsection{Strubbe \& Quataert 2009 (SQ) Model} \label{subsec:SQ09}
We adopt the model developed by \cite{Strubbe&Quataert2009} (hereafter \citetalias{Strubbe&Quataert2009}), who take into account the advection effect. A disk is assumed to form promptly when the debris comes back to the BH, around a circularization radius ($\simeq 2\,R_{\rm T}$). This model describes the accretion flow within the radius where the fallback materials continue to join the disk. The mass accretion rate is the same across the entire disk and set equal to the fallback rate (Eq.~\ref{eq:fallback_rate}), which is justified as long as the accretion (or viscous) timescale is shorter than the fallback time. Including the advection effect allows to capture the transition from the slim to standard disk regimes.

Here, we briefly summarize the main properties of the model and do not repeat the derivation (see \citetalias{Strubbe&Quataert2009} for the details). The scale height and surface density are given by 
\begin{align}
    \label{eq:scale_height}
    H &\simeq 2.1 \times 10^{12} \, {\rm cm} \, 
    g^{-1} \dot{m} M_{\bullet,6}\ , \\
    \label{eq:surface_density}
    \Sigma & \simeq 1.7 \times 10^{3} \, {\rm g} \, {\rm cm}^{-2} \,
    g^{2} \alpha_{-1}^{-1} \dot{m}^{-1} r_{2}^{3/2}\ ,
\end{align}
where $\dot{m} = \dot{M}/\dot{M}_{\rm Edd}$, $r = R/R_{\rm g}$, and $\alpha$ is the viscous parameter \citep{Shakura&Sunyaev1973}. We normalize $r = 100\, r_2$ and $\alpha=0.1 \, \alpha_{-1}$. The factor $g$ accounts for the effect of the advection cooling and is given by 
\begin{align}
    \label{eq:f}
    g &= \frac{1}{2} + \left[\frac{1}{4} + 600\dot{m}^{2} r^{-2}\right]^{1/2}\ .
\end{align}
When the advection cooling is more efficient than the radiation one, the second term in the parenthesis of Eq.~\eqref{eq:f} dominates the first term. Then the factor asymptotes to $g \simeq 24\dot{m} r^{-1}$ representing the slim disk regime. This condition is written by 
\begin{align}
    R \lesssim 49 \, R_{\rm g} \, \dot{m} \ ,
    \label{eq:R_trap}
\end{align}
and the right hand side corresponds to the so-called trapping radius \citep{Begelman1978,Begelman&Meier1982}. For a large radius or a small accretion rate, the condition is violated and $g$ approaches to unity, reducing Eqs.~\eqref{eq:scale_height} and \eqref{eq:surface_density} to the expressions for the standard disk \cite[e.g.,][]{Kato+2008}.

\begin{figure}
    \centering
    \includegraphics[keepaspectratio, scale=0.35,bb=0 0 648 864]{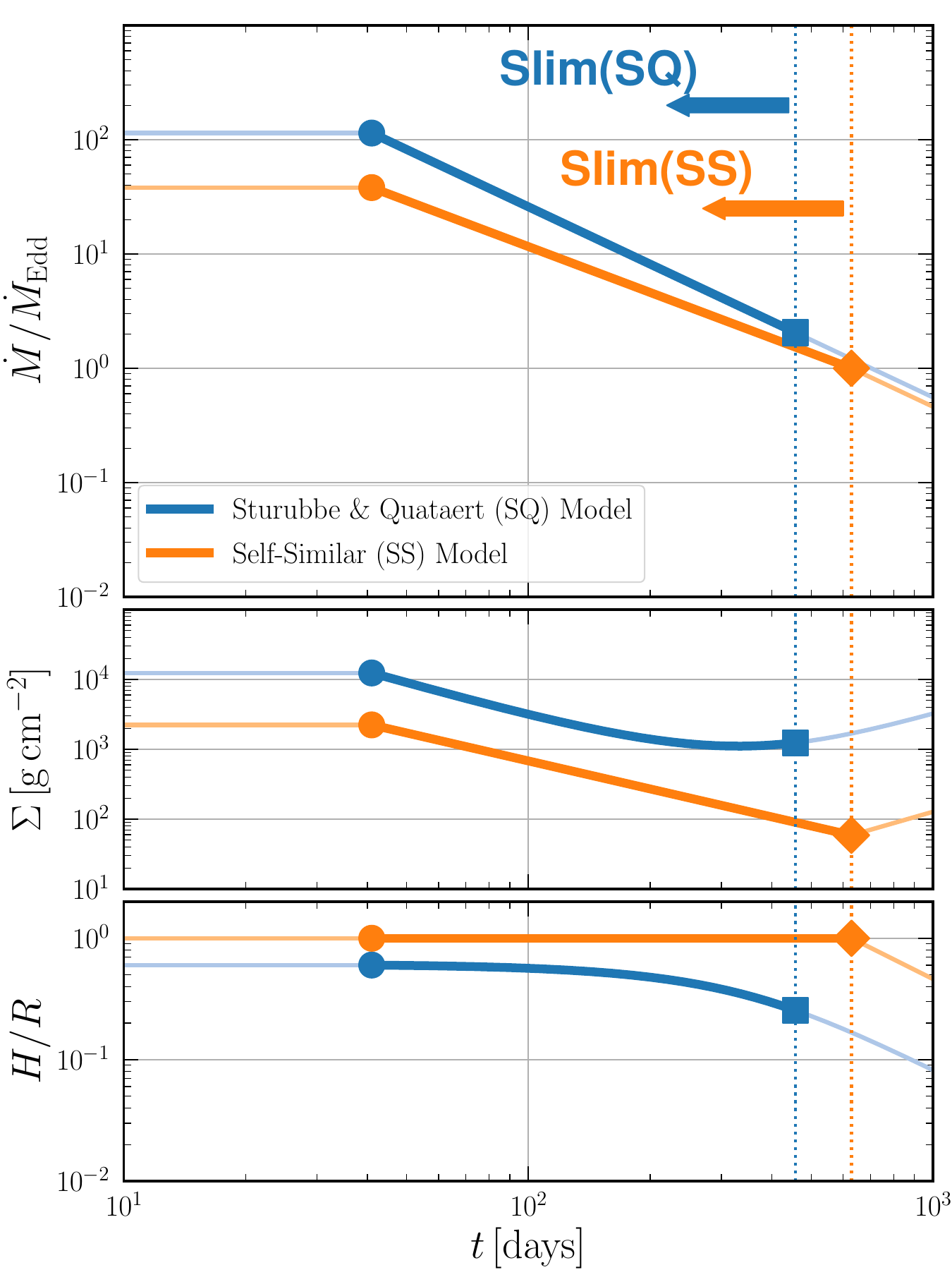}
    \caption{Time evolution of the accretion rate (\textit{upper}), surface density (\textit{middle}), and scale height (\textit{lower}) of the disk for the Strubbe\&Quataert (SQ, blue) and self-similar (SS, orange) models. The parameters are $M_{\bullet} = 10^6 \, M_{\odot}$, $M_{\star,{\rm TDE}} = M_{\odot}$, $R_{\star,{\rm TDE}} = R_{\odot}$, $\alpha=0.1$, and $R = 100 \, R_{\rm g}$. The SQ model is reliable after the fallback time (Eq.~\ref{eq:fallback_time}) and until the photon trapping condition (Eq.~\ref{eq:t_trap}) is violated at square points (or left of the vertical dashed lines). The SS model is valid for $\dot{M} \gtrsim \dot{M}_{\rm Edd}$ and we extrapolate $\dot{M}$ with the same temporal index $\dot{M} \propto t^{-5/3}$ beyond diamond points.}
    \label{fig:Mdot}
\end{figure}

Figure~\ref{fig:Mdot} shows the time evolution of the mass accretion rate, surface density, and scale height calculated for $M_{\bullet} = 10^{6} \, M_{\odot}$, $M_{\star,{\rm TDE}} = M_{\odot}$, $R_{\star,{\rm TDE}} = R_{\odot}$, $\alpha=0.1$, and $R = 100 \, R_{\rm g}$. The evolution is introduced through the time-dependence of the mass accretion rate. During the very early phase of $t \lesssim t_{\rm fb}$, the disk is forming and the accretion rate may not be captured by the fallback rate (Eq.~\ref{eq:fallback_rate}). We simply fix $\dot{M}$ to the same value as the fallback rate at $t = t_{\rm fb}$ in this stage. The accretion rate monotonically declines and becomes comparable to the Eddington rate at
\begin{align}
    \label{eq:t_Edd}
    t_{\rm Edd} = 
    \left(\frac{M_{\star,{\rm TDE}}}{3t_{\rm fb} \dot{M}_{\rm Edd}}\right)^{3/5} t_{\rm fb} 
    \simeq 770 \, {\rm day} \, 
    {\cal M}_{\star,{\rm TDE}}^{1/5} {\cal R}_{\star,{\rm TDE}}^{3/5} M_{\bullet,6}^{-2/5}\ .
\end{align}
At $R = 100 \,R_{\rm g}$, the condition for photon trapping (Eq.~\ref{eq:R_trap}) holds until 
\begin{align}
    t \lesssim t_{\rm pt} 
    &\equiv 0.65 \, r_{2}^{-3/5} t_{\rm Edd} 
    \nonumber \\
    &\simeq 450{\,\rm day\,}
    {\cal M}_{\star,{\rm TDE}}^{1/5}
    {\cal R}_{\star,{\rm TDE}}^{3/5}
    M_{\bullet,6}^{-2/5}
    r_{2}^{-3/5}\ ,
    \label{eq:t_trap}
\end{align}
which is slightly earlier than the above timescale (Eq.~\ref{eq:t_Edd}). During the slim disk state, the surface density decreases and the aspect ratio is constant, but they increase and decrease, respectively, in the standard disk regime.

Note that the increase of $\Sigma$ in the later phase means accumulating mass in the disk, which invalidates our assumption of the quasi stationary disk. This is caused by the longer viscous timescale for lower accretion rate. Indeed this regime is known to be unstable for the viscous and thermal instabilities \citep[e.g.,][]{Lightman&Eardley1974,Shibazaki&Hoshi1975,Pringle1976,Shakura&Sunyaev1976}. As the (externally induced) fallback rate declines, the viscous timescale becomes longer (corresponding to the middle branch of the S-curve in a $\Sigma$-$\dot{M}$ plane, \citealt{Abramowicz+1988}). Therefore, the time evolution in this regime is not reliable while we show it for the illustrative purpose.

\subsection{Self-Similar (SS) Model} 
\label{subsec:self-similar}
In the other model, we consider a self-similar solution of the super-Eddington accretion disk, which is able to capture the viscous spreading. A disk forms promptly in the initial fallback timescale as in the SQ model. Since most of the debris joins the disk in this initial stage, we may regard the disk as isolated and apply a self-similar solution \citep{Pringle1974,Lynden-Bell&Pringle1974,Lyubarskij&Shakura1987,Pringle1991}. Indeed previous studies have shown that the disk asymptotes to a self-similar evolution \citep{Cannizzo+1990,van_Velzen+2019,Alush&Stone2025,Tamilan+2025}. However their solutions are applicable to the late-time evolution with $\dot{M} \lesssim \dot{M}_{\rm Edd}$ and cannot describe the early stage. To capture the nature of super-Eddington disks, we adopt a solution discussed by \cite{Cannizzo&Gehrels2009} in the context of gamma-ray bursts. 

Again, we summarize the basic properties of the solution and defer its detailed derivation to Appendix \ref{appsec:disk}. Under the super-Eddington accretion, the disk cannot cool efficiently and becomes geometrically thick. In particular, we assume the scale height is comparable to the radius:
\begin{align}
    H\simeq R\ .    
\end{align}
The surface density is given by 
\begin{align}
    \label{eq:self_similar_sigma}
    \Sigma(R,t) &= \Sigma_{0} 
    \left(\frac{R}{R_{0}}\right)^{-1/2}
    \left(\frac{t}{t_{0}}\right)^{-4/3} \nonumber \\
    &\quad \times {\rm exp}
    \left[-\frac{1}{9}\left(\frac{R}{R_{0}}\right)^{3/2}
    \left(\frac{t}{t_{0}}\right)^{-1}\right],
\end{align}
where $\Sigma_{0}$, $R_{0}$, and $t_{0}$ are arbitrary normalization of the surface density, radius, and time, respectively. The profile declines exponentially for large radius, which defines a characteristic ``outer edge'' of the viscously expanding disk: $R_{\rm out} \sim R_{0}(t/t_{0})^{2/3}$.

Motivated by the prompt circularization picture, we consider the disk forms over the characteristic fallback timescale of $t_{\rm fb}$ (Eq.~\ref{eq:fallback_time}) and use the solution (Eq.~\ref{eq:self_similar_sigma}) only for $t \geq t_{0}$ by setting $t_{0} = t_{\rm fb}$. Then the normalization radius is automatically determined by the condition for the self-similarity (see Eq.~\ref{appeq:closure_relation}):
\begin{align}
    R_0 &= \left[\frac{\alpha (GM_\bullet)^{1/2} t_{0}}{2}\right]^{2/3}
    \simeq 23 \, R_{\rm T} \, 
    \alpha_{-1}^{2/3} M_{\bullet,6}^{1/3}{\cal M}_{\star,{\rm TDE}}^{-1/3}\ .
\end{align}
Note that this radius is $\sim 10$ times larger than the circularization radius $\simeq 2 \, R_{\rm T}$. However, the viscous timescale corresponding to a radius $R$ is
\begin{align}
    t_{\rm vis}(R) \sim 
    \frac{R^{2}}{\nu_{\rm vis}}
    \sim t_{0}\left(\frac{R}{R_{0}}\right)^{3/2}\ ,
    \label{eq:t_vis}
\end{align}
where $\nu_{\rm vis}$ is the effective viscosity (see Eq.~\ref{appeq:nu}). The timescale becomes comparable to $t_{0}$ for $R=R_{0}$, and we may reasonably assume that over the fallback timescale, the disk can viscously expand to $R_{0}$ satisfying the self-similarity condition. The disk mass is given by
\begin{align}
    \label{eq:disk_mass}
    M_{\rm disk}(t) 
    = \int_{0}^{\infty} 2 \pi R \Sigma(R,t) {\rm d}R 
    = 12 \pi R_{0}^{2} \Sigma_{0} \left(\frac{t}{t_{0}}\right)^{-1/3}\ .
\end{align}
Since about one third of the disrupted stellar mass falls back over $t_{\rm fb}$ (see Eq.~\ref{eq:fallback_rate}) and forms the disk, the normalization of $\Sigma$ is obtained by
\begin{align}
    \Sigma_{0} = \frac{M_{\star,{\rm TDE}}}{36 \pi R_{0}^{2}}\ .
\end{align}
Note we neglected the falling back mass after $t>t_{\rm fb}$, which might be justified because most of the mass joins the disk at the early stage. The time derivative of the disk mass gives the accretion rate:
\begin{align}
    \label{eq:Mdot SS}
    \dot{M} &= 
    \frac{4 \pi R_{0}^{2} \Sigma_{0}}{t_{0}} 
    \left(\frac{t}{t_{0}}\right)^{-4/3}
    = \frac{M_{\star,{\rm TDE}}}{9 t_{\rm fb}}
    \left(\frac{t}{t_{\rm fb}}\right)^{-4/3}\ ,
\end{align}
which is one third of the fallback rate (Eq.~\ref{eq:fallback_rate}) at $t_{\rm fb}$ and declines a little more slowly. 

The time evolution of the mass accretion rate, surface density, and scale height are shown in Fig.~\ref{fig:Mdot}. Since the accretion rate during the disk formation ($t < t_{\rm fb}$) is not reliable, we use the value at $t_{\rm fb}$ and extrapolate it backward as a constant as we did in the SQ model. In addition, at late time the accretion rate becomes smaller than the Eddington one and the solution is no longer valid for
\begin{align}
    t > 
    \left(\frac{M_{\star,{\rm TDE}}}{9 t_{\rm fb} \dot{M}_{\rm Edd}}\right)^{3/4}
    t_{\rm fb}
    \simeq 630 \, {\rm day} \,  
    {\cal R}_{\star,{\rm TDE}}^{3/8}
    {\cal M}_{\star,{\rm TDE}}^{1/2}
    M_{\bullet,6}^{-5/8}\ .
    \label{eq:t_Edd_SS}
\end{align}
While several scenarios for this late phase have been proposed \citep[e.g.,][]{Cannizzo+1990,Shen&Matzner2014,Kaur+2023,Piro&Mockler2025,Alush&Stone2025}, we simply extend the accretion rate by assuming $\dot{M} \propto t^{-k}$ and the radiation-pressure-dominated standard disk. In this regime, the surface density and scale height are given by $\Sigma \propto \dot{M}^{-1} \propto t^{k}$ and $H \propto \dot{M} \propto t^{-k}$, respectively \citep[e.g.,][]{Kato+2008}. Here the temporal index $k$ is not determined a priori, and we consider the conservative case of $k = 5/3$, which means that the accretion rate starts to follow the fallback rate.

\section{QPEs in super-Eddington disk} \label{sec:evolution}
The early-time evolution of QPEs where the EMRI interacts with the slim disk, can be obtained by plunging the expressions of the surface density and scale height into the equations for the QPE observables. While we discuss the results for different two disk models separately, they are qualitatively similar.

\subsection{SQ Model} \label{subsec:evolution_SQ09}
\begin{figure}
    \centering
    \includegraphics[keepaspectratio, scale=0.35,bb=0 0 648 648]{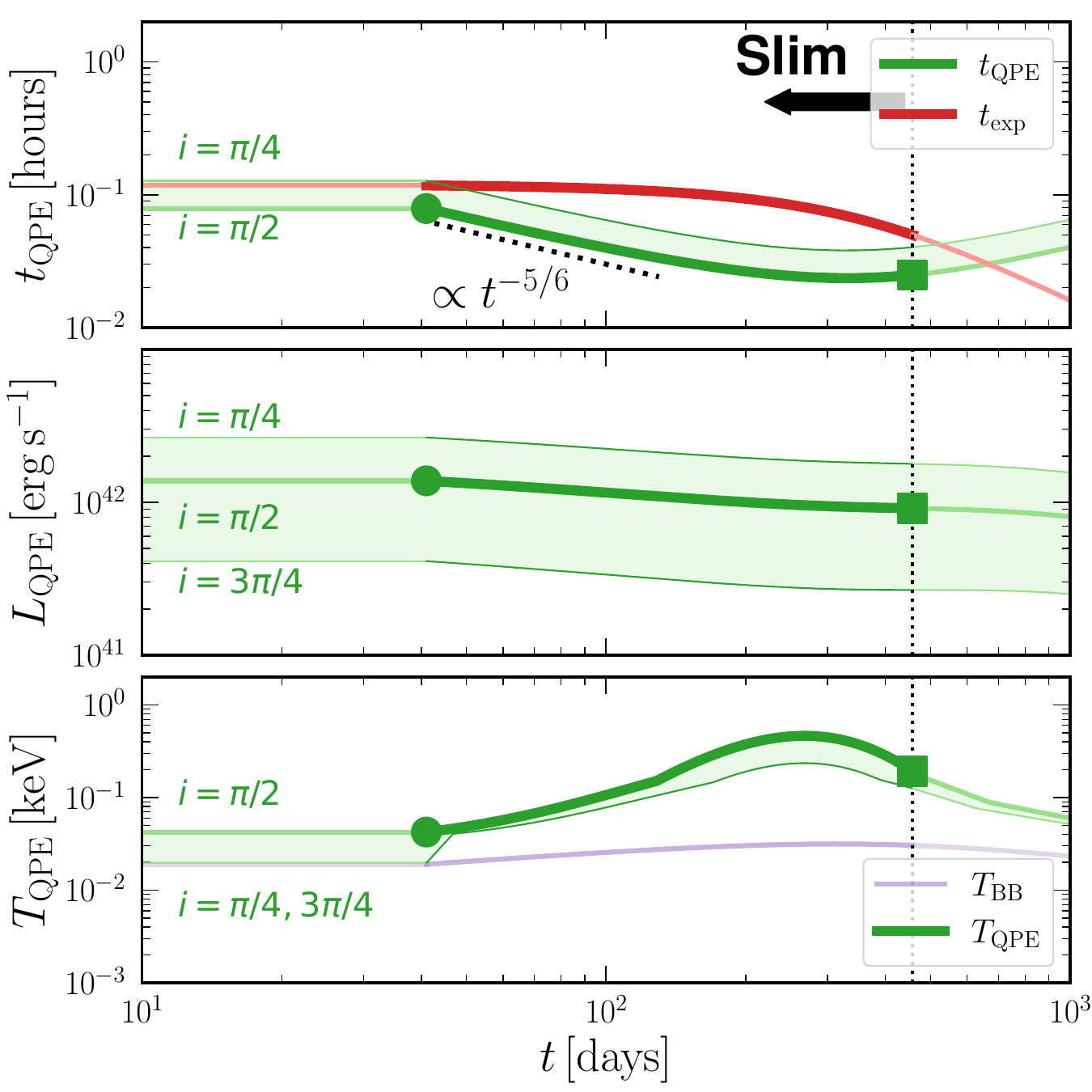}
    \caption{Time evolution of the QPE duration (\textit{upper}), luminosity (\textit{middle}), and characteristic temperature (\textit{lower}) for the SQ model. The disk parameters are the same as Fig.~\ref{fig:Mdot}, and the EMRI parameters are $R_{\star} = R_{\odot}$ and $i = \pi/2$. The photon trapping condition (Eq.~\ref{eq:t_trap}) is satisfied to the left of the vertical dashed lines. In the upper panel, the expansion time is also plotted by the red line. The shaded regions around the thick green curves show the possible variation due to different inclination angles. The angles corresponding to the boundaries are shown explicitly. Dotted lines denote the scaling of the quantities in the slim-disk phase (see Eq.~\ref{eq:t_QPE_SQ}). During the almost entire evolution, the temperature is larger than the blackbody one ($T_{\rm BB}$, purple) due to the insufficient thermalization.}
    \label{fig:slim}
\end{figure}

Figure~\ref{fig:slim} depicts the time evolution of the QPE duration, luminosity, and characteristic temperature for the SQ model. They are obtained by using the disk properties in Fig.~\ref{fig:Mdot} with parameters of $M_{\bullet} = 10^{6} \, M_{\odot}$, $M_{\star,{\rm TDE}} = M_{\odot}$, $R_{\star} = R_{\star,{\rm TDE}} = R_{\odot}$, $a (= R) = 100 \, R_{\rm g}$, $i = \pi/2$ (a perpendicular orbit to the disk), and $\alpha = 0.1$. It should be noted that at the early stage of $t < t_{\rm fb}$, the disk and hence QPE properties do not change over time because we artificially fix the accretion rate.

The short duration (or small duty cycle) and high temperature characterize the slim-disk QPEs. In the slim-disk state, the QPE duration and temperature decrease and increase over time, respectively, while the luminosity remains nearly constant. This is because the surface density and ejecta mass monotonically decrease over time for the slim-disk phase (see the middle panel of Fig.~\ref{fig:Mdot}). Smaller ejecta mass results in shorter diffusion time and less efficient thermalization. The luminosity is independent of the mass (see \citetalias{Linial&Metzger2023}). This evolution continues until the condition for the photon-trapping becomes invalidated at $\simeq 450\,{\rm day}$ (Eq.~\ref{eq:t_trap}) and the disk transitions to the standard-disk state. The duration and temperature have the minimum and maximum there, respectively.

Quantitative expressions for the QPE duration and luminosity in the slim-disk phase are obtained by substituting Eqs.~\eqref{eq:scale_height} and \eqref{eq:surface_density} to Eqs.~\eqref{eq:duration}, \eqref{eq:luminosity}, and taking a limit of $g \simeq 24 \dot{m} r^{-1}$ (see Eq.~\ref{eq:f}):
\begin{align}
    t_{\rm QPE} &\simeq 9.3 \times 10^{-2} \, {\rm hr} \,
    \frac{1}{(\sin{i}\sin{(i/2)})^{1/2}} 
    \nonumber \\
    &\quad \times 
    \frac{{\cal R}_{\star} {\cal M}_{\star,{\rm TDE}}}
    {\alpha_{-1}^{1/2} M_{\bullet,6}^{3/4} {\cal R}_{\star,{\rm TDE}}^{3/4}}
    \left(\frac{t}{t_{\rm fb}}\right)^{-5/6}\ ,
    \label{eq:t_QPE_SQ} \\
    L_{\rm QPE} &\simeq 9.1 \times 10^{41} \, {\rm erg} \, {\rm s}^{-1} \,
    \frac{\sin^{2}{(i/2)}}{\sin^{1/3}{i}}
    \frac{{\cal R}_{\star}^{2/3} M_{\bullet,6}^{1/3}}{a_{2}^{2/3}}\ .
    \label{eq:L_QPE_SQ}
\end{align}
The minimum duration is given by
\begin{align}
    t_{{\rm QPE,min}} &\simeq 2.6 \times 10^{-2} \, {\rm hr} \,
    \frac{1}{(\sin{i}\sin{(i/2)})^{1/2}}
    \frac{{\cal R}_{\star} a_{2}^{1/2}}{\alpha_{-1}^{1/2}}\ .
\end{align}

\begin{figure}
    \centering
    \includegraphics[keepaspectratio, scale=0.4,bb=0 0 576 432]{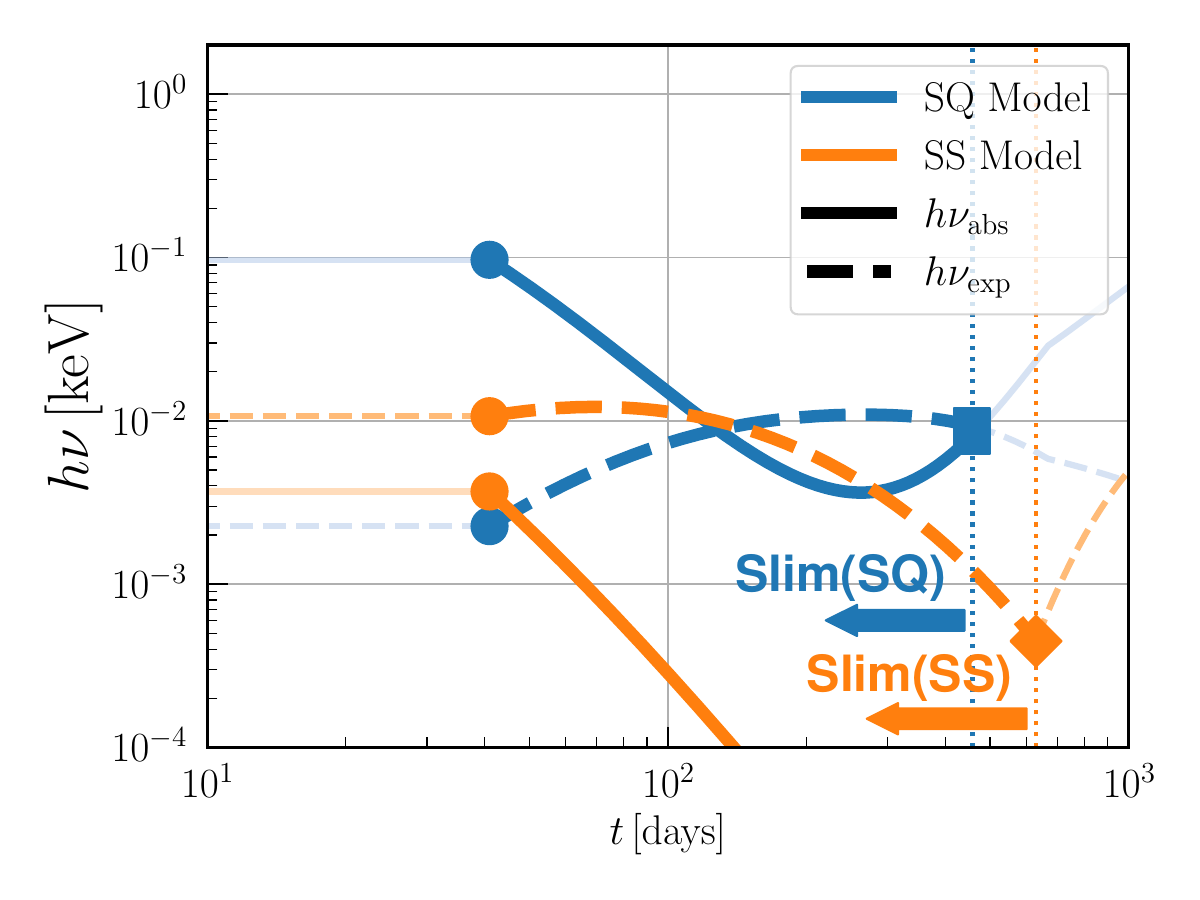}
    \caption{
    Time evolution of the critical photon energies for Comptonization. The solid and dashed curves show the photon energies above which photons can be upscattered without free-free absorption, and within the expansion timescale, respectively. The larger of the two determines the contribution of Comptonization. The blue and orange curves represent the SQ (blue) and SS (orange) models, respectively.}
    \label{fig:hnu}
\end{figure}

The derivation of the scaling for the temperature evolution is not as straightforward as for the other quantities, because of the contribution of Comptonization, which enters the calculation through the more complicated form of Eq.~\eqref{eq:temperature_comptonization}. Instead of providing an explicit scaling relation, we just describe the temporal behavior. Figure~\ref{fig:hnu} shows the critical photon energies determining the importance of Comptonization, $\nu_{\rm exp}$ and $\nu_{\rm abs}$. Initially, the minimum energy is set by the free-free absorption, $\nu_{\rm abs} > \nu_{\rm exp}$, due to the high density. As the ejecta expands, the absorption becomes less efficient and eventually the minimum energy is set by $\nu_{\rm exp}$ at $\simeq 120 \, {\rm day}$. During the evolution, the parameter for Comptonization does not change significantly and has a value of $\xi \simeq 5$.

Different inclination angles from $i = \pi/2$ do not significantly impact the overall behavior of the QPE quantities. The shaded regions in Fig.~\ref{fig:slim} show the possible range of the observables for different angles ($\pi/4 \leq i \leq 3\pi/4$). However, this insensitivity is due to the relatively small range of the inclination angle imposed by the condition to produce QPEs in thick disks (see Fig.~\ref{fig:inc_vel_num}). For the thin disk, the range is wider and its impact is probably more important.

\subsection{SS Model} \label{subsec:evolution_SS}
\begin{figure}
    \centering
    \includegraphics[keepaspectratio, scale=0.35,bb=0 0 648 648]{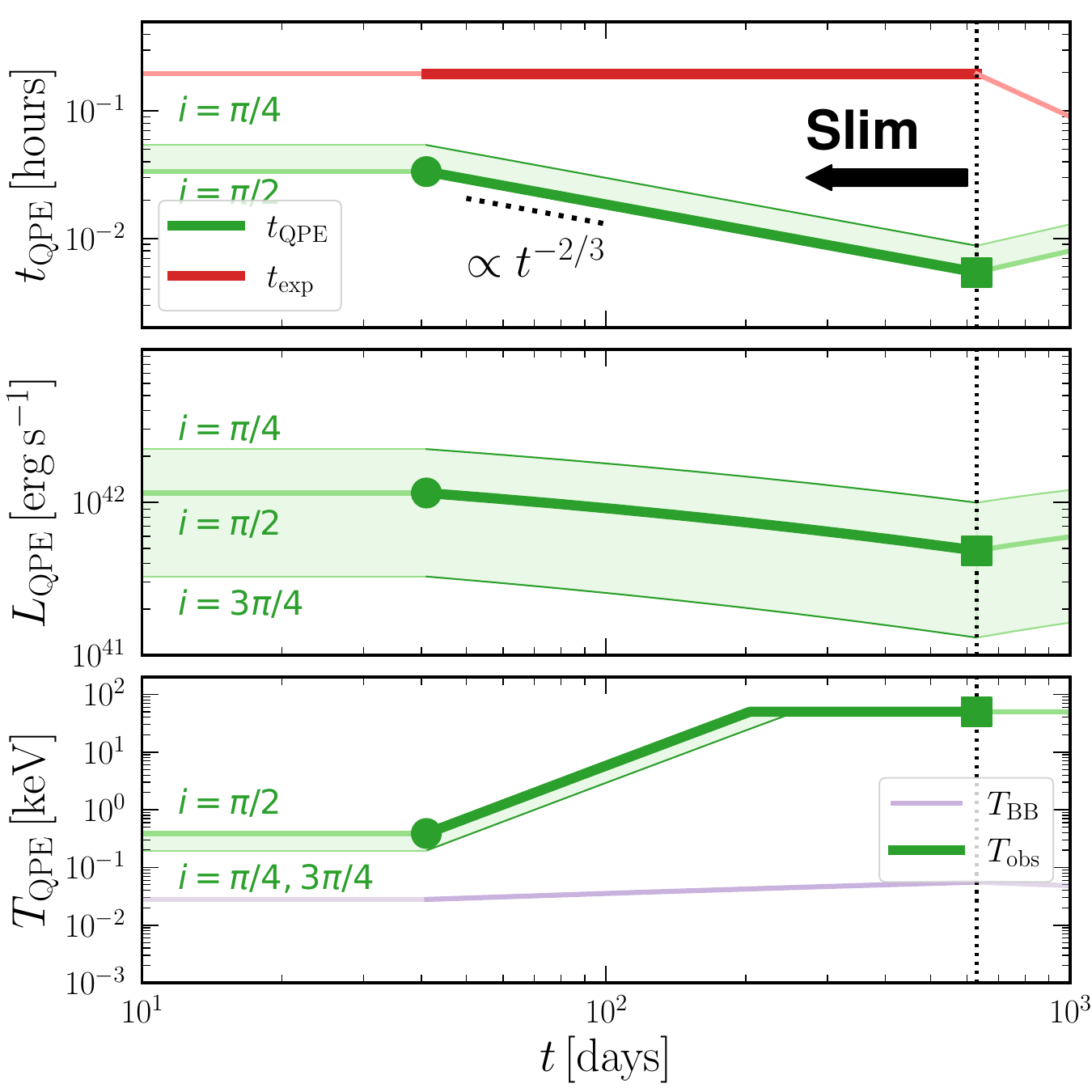}
    \caption{The same as Fig.~\ref{fig:slim} but for the SS model. The temperature is capped by the theoretical upper limit by the pair creation, $T_{\rm QPE} \lesssim 50\,{\rm keV}$.}
    \label{fig:ss}
\end{figure}

Figure~\ref{fig:ss} shows the results for the SS model whose disk properties are shown in Fig.~\ref{fig:Mdot}. The QPE evolution is qualitatively similar to the SQ model because of the similar disk evolution, but a relatively lower surface density results in a shorter duration and higher temperature. We remark that the self-similar solution is valid only for $\dot{M} \gtrsim \dot{M}_{\rm Edd}$ (see Eq.~\ref{eq:t_Edd_SS} for the corresponding time). The QPE duration and luminosity are derived in the same way as the SQ model: 
\begin{align}
    \label{eq:t_QPE_SS}
    t_{\rm QPE} &\simeq 4.0 \times 10^{-2} \, {\rm hr} \,
    \frac{1}{(\sin{i}\sin{(i/2)})^{1/2}} \nonumber \\
    &\quad \times 
    \frac{{\cal R}_{\star}{\cal M}_{\star,{\rm TDE}}}
    {{M_{\bullet,6}^{3/4}}\alpha_{-1}^{1/2}{\cal R}_{\star,{\rm TDE}}^{3/4}}
    \left(\frac{t}{t_{\rm fb}}\right)^{-2/3}\ ,\\
    \label{eq:L_QPE_SS}
    L_{\rm QPE} &\simeq 1.1 \times 10^{42} \, {\rm erg} \, {\rm s}^{-1} \,
    \frac{\sin^{2}{(i/2)}}{\sin^{1/3}{i}}
    \frac{{\cal R}_{\star}^{2/3}M_{\bullet,6}^{1/3}}{a_{2}^{2/3}}\ .
\end{align}
The QPE duration has the minimum around the transition: 
\begin{align}
    t_{\rm QPE,min} &\simeq 6.5 \times 10^{-3} \, {\rm hr} \,
    \frac{1}{(\sin{i}\sin{(i/2)})^{1/2}}
    \frac{{\cal R}_{\star}}{\alpha_{-1}^{1/2}}\ .
\end{align}
The temperature is higher than that of the SQ model, and in particular it reaches the upper bound of $\sim 50\,{\rm keV}$. The Comptonization play a role to determines the temperature, but the minimum frequency is always set by the expansion, $\nu_{\rm exp} > \nu_{\rm abs}$ because of the low density.

The disk enters the standard-disk state at around the timescale of Eq.~\eqref{eq:t_Edd_SS}. The following evolution of QPEs has been studied in the previous works \citep[e.g.,][]{Linial&Metzger2023, Franchini+2023, Tagawa&Haiman2023}, and it is not the focus of our study. However, for the illustrative purpose, we extend the lines by calculating the evolution by using the standard disk properties.

\section{Discussions} \label{sec:discussion}
\subsection{The QPE evolution for different parameters} 
\begin{figure*}
    \centering
    \includegraphics[keepaspectratio, scale=0.3,bb=0 0 1728 1296]{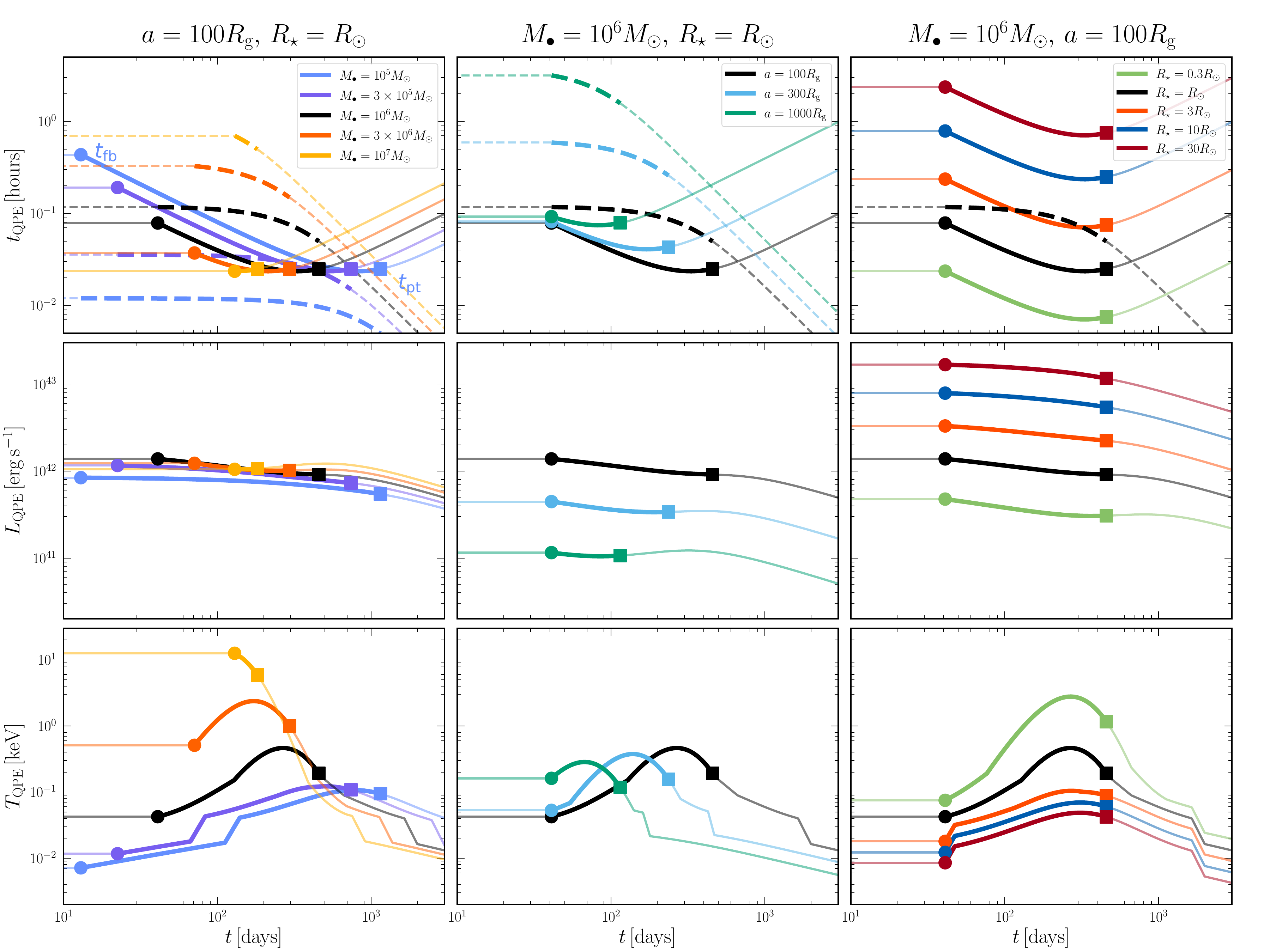}
    \caption{The QPE evolution for varied BH masses (left panels), locations of the EMRI (middle), and sizes of the EMRI (right) in the SQ model. The results calculated for our fiducial parameters ($M_{\bullet}=10^{6} \, M_{\odot}$, $a = 100 \, R_{\rm g}$, and $R_{\star} = R_{\odot}$, as used in Fig.~\ref{fig:slim}) are shown by black curves. On each curves, the moment of $t = t_{\rm fb}$ and $t = t_{\rm pt}$ are shown by circles and squares. In the upper panels, the expansion time $t_{\rm exp}$ is also plotted by the dashed lines.}
    \label{fig:parameter}
\end{figure*}

The QPE properties and their evolution are determined by multiple parameters, such as the BH mass $M_{\bullet}$, radius $R_{\star}$ and location $a$ of the EMRI, and mass $M_{\star,{\rm TDE}}$ and radius $R_{\star,{\rm TDE}}$ of the disrupted star as well as the viscosity parameter $\alpha$. While the parameter dependence of the key observables is derived in the previous section, we discuss how the results are altered in particular for different BH mass, EMRI location, and radius. Figure~\ref{fig:parameter} shows the QPE evolution calculated for different parameters from our fiducial values in the SQ model. The corresponding results for the SS model are presented in Appendix~\ref{appsec:parameter_ss}. 

Varying the BH mass (left column) impacts the QPE signals through changing the disk properties for the fixed location of $a = 100 \, R_{\rm g}$. A Larger BH mass increases the spatial scale (via $R_{\rm g} \propto M_{\bullet}$), which lengthens $t_{\rm fb}$ (circles, Eq.~\ref{eq:fallback_time}) and shortens $t_{\rm pt}$ (squares, Eq.~\ref{eq:t_trap}), thereby resulting in a shorter slim-disk phase. Moreover, the smaller surface density and hence the ejecta mass decrease the QPE duration (e.g., Eq.~\ref{eq:t_QPE_SQ}), but increase the temperature. The luminosity is relatively insensitive to the BH mass. The QPE recurrence time also decreases as $P_{\rm QPE} \propto M_{\bullet}^{-1/2}$ (Eq.~\ref{eq:P_QPE}), and the slightly sensitive dependence of $t_{\rm QPE} \propto M_{\bullet}^{-3/4}$ results in a decreasing duty cycle $\propto M_{\bullet}^{-1/4}$. We find that for $M_{\bullet} \gtrsim 10^{6} \, M_{\odot}$, the ejecta's expansion timescale is shorter than the diffusion time; that is, QPEs occur immediately after the EMRI emerges from the disk.

The qualitative behavior of the QPE temperature differs from that of the fiducial case for different BH masses. For $M_{\bullet} > 10^{6} \, M_{\odot}$, the radiation is initially out of thermal equilibrium, and the temperature deviates from the blackbody value. This is simply because of the low surface density for these BH masses, and an insufficient free-free absorption $\nu_{\rm abs} < \nu_{\rm exp}$. For lower BH mass $M_{\bullet} \lesssim 3 \times 10^{5} \, M_{\odot}$, the thermalization is initially efficient and the temperature coincides with the blackbody one. As  the surface density decreases, the temperature increases. Comptonization starts to contribute in the temperature increase at $\sim 100\,{\rm days}$ causing a small jump in the evolution. This behavior may be artificially produced by our simple prescription for including Componitzation, and realistically the evolution should be smoother.

Note that for $M_{\bullet} \gtrsim 10^{6} \, M_{\odot}$, the circularization radius is smaller than the EMRI location, $2 R_{\rm T} \lesssim a = 100 \, R_{\rm g}$ (see Eq.~\ref{eq:tidal_radius}), invalidating the SQ disk model. However, if the disk spreads via viscosity as in the SS model, it probably expands to the location of the EMRI in a short timescale, $t_{\rm vis}(100\,R_{\rm g}) \simeq 0.026 \, t_{\rm fb} M_{\bullet,6}^{1/2} r_{2}^{3/2}$ (Eq.~\ref{eq:t_vis}).

In the middle column, we increase the distance of the EMRI from the BH (note that the distance is bounded below by the tidal radius, $a \gtrsim R_{\rm T} \simeq 47 \, R_{\rm g}$, Eq.~\ref{eq:tidal_radius}). At larger distance, in addition to the smaller orbital velocity, the scale height increases and the surface density decreases in the slim-disk phase. The reduction in both the velocity and surface density cancels out and gives the same QPE duration. A smaller ejecta energy gives a lower QPE luminosity. 

Finally, varying the EMRI's radius (right column) does not change the functional form of each quantity but only the normalization for the duration and luminosity. This is easily understood by noting the larger radius gives larger ejecta mass, which results in longer diffusion time and higher luminosity. The temperature evolution, however, shows subtle differences. For $R_{\star} > 3 \, R_{\odot}$, the temperature is determined by Comptonization with the minimum energy set by free-free absorption. For $R_{\star} = 0.3 \, R_{\odot}$, the evolution closely follows that of the fiducial case. 

\subsection{Detectability of the slim-disk QPEs}
\begin{figure}
    \centering
    \includegraphics[keepaspectratio, scale=0.4,bb=0 0 576 432]{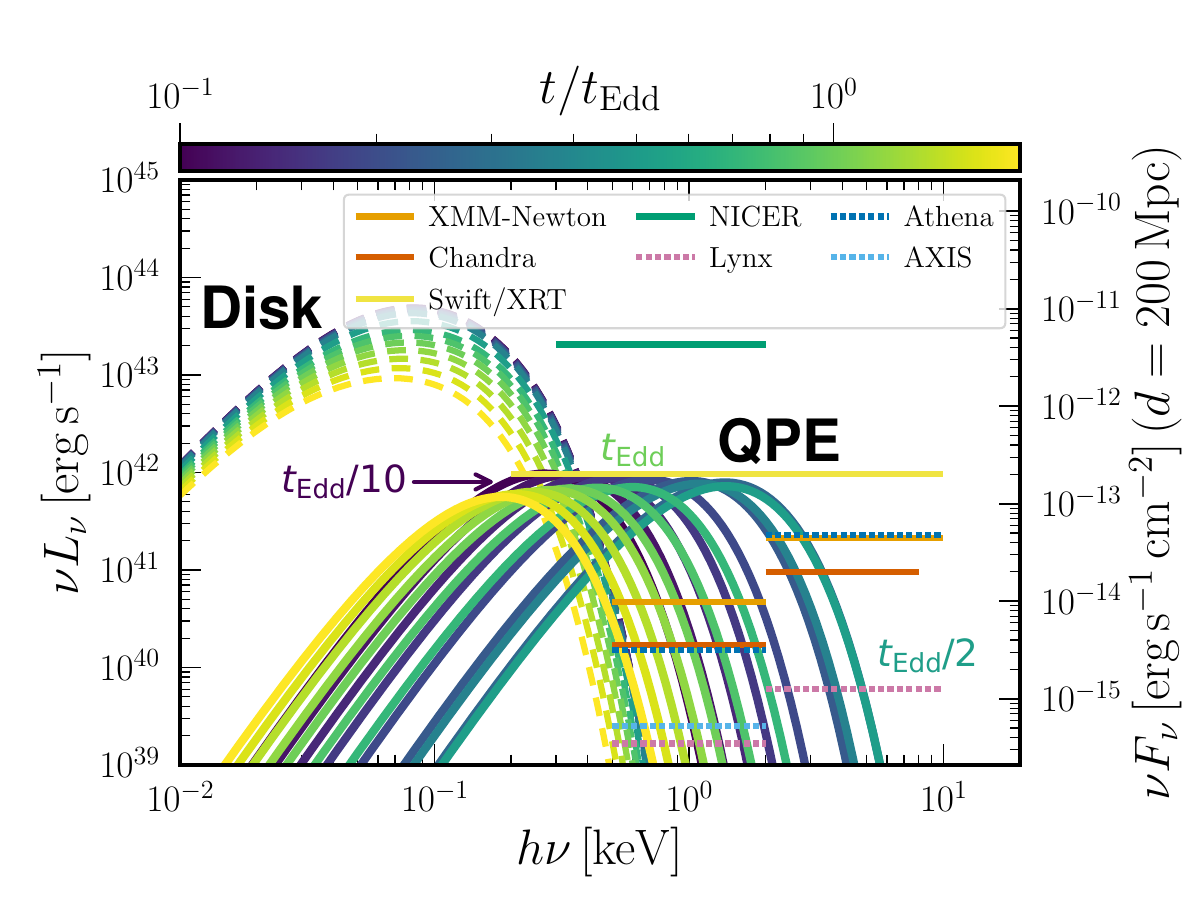}
    \caption{The SEDs of the QPE (solid) and disk (dashed) components at different times for the SQ model. The parameters are the same as Fig.~\ref{fig:slim}. The right vertical axis shows the flux for a QPE at $d = 200 \, {\rm Mpc}$. Horizontal lines denote the sensitivities of various X-ray observatories for the exposure times of $100 \, {\rm s}$ (see also Table~\ref{table:sensitivity}).}
    \label{fig:QPE_SED}
\end{figure}

QPEs in slim disk are characterized by high temperatures $\simeq 0.5-50 \, {\rm keV}$, which makes them distinct from the (persistent) disk emission and detectable by the current and future X-ray telescopes. The slim-disk luminosity may be comparable to the Eddington one $L_{\rm Edd} \sim 10^{44}{\,\rm erg \, s^{-1}\,} M_{\bullet,6}$ \citep[e.g.,][]{Watarai&Fukue1999}, much larger than the QPE luminosity. However, as for the currently detected QPEs, the disk temperature is lower than the QPE one, and the QPE emission is not buried in the quiescent disk emission. To demonstrate this, we show the evolution of the spectral energy distributions (SEDs) of both disk and QPE components in Fig.~\ref{fig:QPE_SED}. The disk SED is calculated by integrating the blackbody spectrum contributed from each annulus over the disk \cite[e.g.,][]{Kato+2008}:
\begin{align}
    (\nu L_{\nu})_{\rm disk} 
    &= \int_{R_{\rm in}}^{R_{\rm out}} \nu \pi B_{\nu}(T_{\rm eff}(R)) 
    2 \pi R {\rm d}R\ ,
    \label{eq:SED_disk}
\end{align}
where $B_{\nu}(T)$ is the Planck function, $T_{\rm eff}(R)$ is the effective temperature (see Eq.~19 in \citetalias{Strubbe&Quataert2009}), and $R_{\rm in(out)}$ is the inner (outer) edge of the disk. We set $R_{\rm in} = 6\,R_{\rm g}$ and $R_{\rm out} = 2\,R_{\rm T}$ in the SQ model. In particular the effective temperature has a maximum 
\begin{align}
    T_{\rm eff,max} \simeq 2.8 \times 10^{-2} \, {\rm keV} \, 
    M_{\bullet,6}^{-1/4}\ ,
\end{align}
at $R \simeq 1.6 \, R_{\rm in}$. Here we use $g \sim 24\dot{m}r^{-1}$.
The QPE SED is obtained by using the Wien spectrum: $(\nu L_{\nu})_{\rm QPE} \propto \nu^{3} \exp(-h\nu/k_{\rm B}T_{\rm QPE})$. The normalization is determined by requiring that the total luminosity coincides with the QPE luminosity.

We remark that a realistic QPE SED is likely to exhibit a more complex shape than that considered here, for the following two reasons. First, the spectrum below the Wien peak may be shallower because free-free emission, rather than saturated Comptonization, dominates the SED for mild Comptonization $1 \lesssim \xi \lesssim 10$. This does not, however, change our conclusion that the QPE component is discernible from the disk emission. Second, a corona may play a role. As seen in active galactic nuclei, if hot and tenuous plasma (a so-called ``corona'') surrounds the disk, it can Comptonize both the QPE and disk emissions, thereby modifying the observed spectra relative to those assumed here. Although current observations have begun to reveal possible signatures of such coronae \citep[e.g.,][]{Guolo+2024}, the sample remains limited. Therefore, we neglect the effects of a corona in this work.

As shown in Fig.~\ref{fig:QPE_SED}, the QPE emission is initially buried in the disk emission, but as the QPE temperature increases to $\gtrsim \, {\rm keV}$, the QPE appears above the Wien tail of the disk emission. After the state transition from the slim to standard disks at $\simeq t_{\rm pt}(\sim 0.5 \, t_{\rm Edd})$, the temperature starts to decrease and becomes lower than the disk temperature.

To check the detectability of the QPEs, we translate the luminosity into the flux assuming a distance of $d = 200\,{\rm Mpc}$ ($z \simeq 0.05$) in the right vertical axis of Fig.~\ref{fig:QPE_SED}. Owing to their high temperatures $\gtrsim 0.5-1.0 \, {\rm keV}$, the slim-disk QPEs can be targets for both soft and hard X-ray telescopes such as \textit{XMM-Newton}, \textit{Chandra}, \textit{Swift/XRT}, and \textit{NICER}. In Fig.~\ref{fig:QPE_SED}, we draw the sensitivities of these telescopes, assuming an exposure time of $100 \, {\rm s}$. These sensitivities are derived from reference values listed in Table~\ref{table:sensitivity} and are scaled under the assumption that sensitivity is reduced with the inverse square root of exposure time. While \textit{Chandra} and \textit{XMM-Newton} can detect the early QPE signals, \textit{Swift/XRT} and \textit{NICER} might not have the required capabilities. In addition, future X-ray observations may be able to detect more distant QPEs. For example, \textit{AXIS}, whose expected sensitivity (see Table~\ref{table:sensitivity}) is $\sim 5.29 \times 10^{-16} \, {\rm erg} \, {\rm cm}^{-2} \, {\rm s}^{-1}$ under the same exposure-scaling assumption, can detect QPEs up to $d \lesssim 4.0 \, {\rm Gpc}$. 

\begin{table*}
    \caption{Sensitivities of the current and future X-ray telescopes.}
    \label{table:sensitivity}
    \begin{center}
    \scalebox{1.0}{
    \begin{tabular}{ccccc}
        \hline
        Name & Band [keV] & Sensitivity $[{\rm erg}\,{\rm s}^{-1}\,{\rm cm}^{-2}]$ & Exposure time [s] & Ref. \\
        \hline\hline
        XMM-Newton & 0.5-2.0 & $3.1\times10^{-16}$ & $10^{5}$ & \citet{Hasinger+2001} \\
                   & 2.0-10 & $1.4\times10^{-15}$ & $10^{5}$ & \citet{Hasinger+2001} \\
        Chandra & 0.5-2.0 & $2.5\times10^{-17}$ & $2\times10^{6}$ & \citet{Alexander+2003} \\
                & 2.0-8.0 & $1.4\times10^{-16}$ & $2\times10^{6}$ & \citet{Alexander+2003} \\ 
        Swift/XRT & 0.2-10 & $2\times10^{-14}$ & $10^{4}$ & \citet{Burrows+2005} \\
        NICER & 0.3-2.0 & $4.3\times10^{-13}$ & $10^{4}$ & \citet{Ramillard+2022} \\
        \hline
        Lynx & 0.5-2.0 & $1.10\times10^{-16}$ & $10^{3}$ & \citet{Lops+2023} \\
        & 2.0-10 & $3.98\times10^{-16}$ & $10^{3}$ & \citet{Lops+2023} \\
        Athena & 0.5-2.0 & $1.00\times10^{-16}$ & $10^{5}$ & \citet{Lops+2023} \\
        & 2.0-10 & $1.50\times10^{-15}$ & $10^{5}$ & \citet{Lops+2023} \\
        AXIS & 0.5-2.0 & $2\times10^{-18}$ & $7\times10^{6}$ & \citet{Cappelluti+2024}
        \\
        \hline
    \end{tabular}
    }
    \end{center}
\end{table*}

\subsection{Effects of disk wind}
A super-Eddington disk tends to launch strong outflows \citep[e.g.,][]{Ohsuga+2005,Sadowski+2016,Jiang+2019,Pacucci&Narayan2024}. While we have neglected such outflows in our calculations, powerful disk winds can influence our results by altering the disk profile, blocking the QPE emission, and potentially modifying the dynamics of the ejecta. First, the presence of winds changes the disk profile, in particular reducing the surface density, which generally results in hotter and shorter QPEs. This effect could be incorporated by adopting the disk model with one including the wind-driven mass loss.

Second, the wind is optically thick and inevitably obscures the inner region of the disk. Time variability shorter than the diffusion timescale from the wind,
\begin{align}
    t_{\rm diff,wind} 
    = \frac{\kappa_{\rm T} \dot{M}_{\rm w}}{4 \pi c v_{\rm w}}
    \simeq 1.4{\,\rm hr\,} \dot{m}_{\rm w} \left(\frac{v_{\rm w}}{0.01\,c}\right)^{-1}\ ,
\end{align}
is washed out, where $\dot{M}_{\rm w}=\dot{m}_{\rm w}\dot{M}_{\rm Edd}$ and $v_{\rm w}$ are the wind mass-loss rate and velocity, respectively. This timescale is longer than the typical duration of the slim-disk QPEs (Eq.~\ref{eq:duration}) unless the EMRI has a large radius of $R_{\star} \gtrsim R_{\odot}$ (see Fig.~\ref{fig:parameter}). However, we note that roughly half of optically-discovered TDEs exhibit X-ray emissions \citep[e.g.,][]{Gezari+2017,Guolo+2024}. In the rapid-circularization scenario, an optically thick wind is driven and reprocesses the X-rays from the disk to optical photons (\citealt{Metzger&Stone2016,Dai+2018,Lu&Bonnerot2020}, but see \citealt{Piran+2015,Matsumoto&Piran2021}). Within this framework, X-rays observed alongside optical emissions imply that the disk wind does not completely cover the inner region, but leaves a ``hole'' through which the X-rays can escape. Indeed numerical simulations show that the outflow has a funnel along the vertical direction of the disk, and could explain the diversity of X-ray properties of optical TDEs \citep{Dai+2018,Thomsen+2022}. Therefore, the wind only reduces the fraction of detectable QPEs, but we may still expect that they can be detected from the polar direction. The fraction of detectable QPEs can be estimated by comparing the solid angle subtended by the funnel, $\simeq1-\cos\theta$, where $\theta$ is the half opening angle of the funnel. For a value of $\theta\simeq\pi/4$, as suggested by simulations \citep{Dai+2018}, we naively expect that roughly $\sim30\,\%$ of optical TDEs accompanied by late-time ``normal'' QPEs should also exhibit early-time QPEs.

Finally, the powerful disk wind probably pushes back the QPE ejecta and hinders its breakout. Whether such suppression occurs or not depends on the relevant timescales. As discussed above, for smaller $M_{\bullet}$ the QPEs are produced before significant ejecta expansion, allowing us to reasonably neglect the wind. For larger $M_\bullet$, however, QPEs occur only after the ejecta has expanded, and in this latter case we need to consider the conditions under which the ejecta can expand against the disk wind and still produce QPEs. For the disk wind, we assume that the mass-loss rate depends on radius as $\dot{M}_{\rm w}(R)\propto R^p$ \citep[e.g.,][]{Blandford&Begelman1999}, where the power-law index typically lies in the range $0\lesssim p \lesssim 1$, and we set $p=0.5$. Since the wind velocity is reasonably given by the escape speed, the total wind energy is dominated by material launched from the inner region. The kinetic luminosity is given by 
\begin{align}
\dot{E}_{\rm w}\simeq \dot{M}_{\rm w}(R_{\rm in})\frac{GM_\bullet}{R_{\rm in}}=\dot{M}_{\rm w}(R)\left(\frac{R}{R_{\rm in}}\right)^p\frac{GM_\bullet}{R_{\rm in}}\ .
\end{align}
In the second equality we use the radial scaling of $\dot{M}_{\rm w}$. The ejecta energy and diffusion timescale (or equivalently the QPE duration) are estimated from Eqs.~\eqref{eq:ejecta_energy} and \eqref{eq:duration}:
\begin{align}
E_{\rm ej}\simeq\frac{R_\star^2v_{\rm K}\dot{M}_{\rm w}(R)}{4\alpha R}{\,\,\,\text{and}\,\,\,}t_{\rm QPE}\simeq\left(\frac{\kappa_{\rm T}R_\star^2\dot{M}_{\rm w}(R)}{4\pi \alpha c R v_{\rm K}^2}\right)^{1/2}\ ,
\end{align}
where we neglect the inclination-angle dependence, and obtain the surface density by using $\dot{M}\sim\dot{M}_{\rm w}(R)\sim 2\pi R v_r\Sigma$. The radial velocity is roughly given by $v_r\sim\alpha v_{\rm K}$ in the thick disk. For an isotropic disk wind, the fraction of the wind material interacting with the QPE ejecta may be estimated from the ratio of solid angles subtended by the wind and ejecta: $\sim(v_{\rm K}t_{\rm QPE}/R)^2$. Therefore, the wind energy available to push the ejecta during a single QPE emission episode is 
\begin{align}
\frac{E_{\rm w}}{E_{\rm ej}}&\simeq\frac{\dot{E}_{\rm w}t_{\rm QPE}(v_{\rm K}t_{\rm QPE}/R)^2}{E_{\rm ej}}
    \nonumber\\
&\simeq\frac{4[\dot{m}_{\rm w}(R)]^{3/2}}{\alpha^{1/2}\epsilon^{3/2}}\left(\frac{R}{R_{\rm g}}\right)^{p-3/2}\left(\frac{R_{\rm in}}{R_{\rm g}}\right)^{-p-1}\left(\frac{R_\star}{R_{\rm g}}\right)
    \nonumber\\
&\simeq0.13\,\alpha_{-1}^{-1/2}\epsilon_{-1}^{-3/2}r_2^{p-3/2}[\dot{m}_{\rm w}(R)]^{3/2}{\cal R}_{\star}M_{\bullet,6}^{-1}\ .
\end{align}
When the wind mass-loss rate slightly exceeds the Eddington rate, $\dot{M}_{\rm w}(R)\gtrsim4\,\dot{M}_{\rm Edd}$, the QPE ejecta can no longer expand freely, and the QPE emission should be suppressed.

In summary, we find that roughly $30 \, \%$ of TDEs which are accompanied by late-time (normal) QPEs can also show early-time QPEs through a tenuous funnel region. However, the QPEs with durations longer than the expansion timescale would emerge only around the end of the super-Eddington phase. Currently no QPE-like emissions have been detected in X-ray follow-ups of optical TDEs within one year after the optical peak, although the sensitivities of current X-ray observatories are sufficient to them (see Fig.~\ref{fig:QPE_SED}). This may be natural because the duty cycle of the slim-disk QPE, $\sim 10^{-2}$ is much smaller than that of the standard disk $\sim 10^{-1}$, which could be even smaller when a powerful disk wind presents, making the detection of the QPE signal challenging. Actually typical observation time of X-ray follow-ups is $\sim 10^{3-4} \, {\rm s}$ \citep[e.g.,][]{Holoien+2016,Gezari+2017}, not entirely covering the recurrence timescale. In addition, even if a follow-up observation happened to coincide with single QPE burst, the signal could still be obscured by a super-Eddington wind as discussed above. To facilitate the discovery of high-temperature QPEs, longer-duration X-ray monitoring of a larger sample of TDEs is necessary. A non-detection in such an intensive follow-up campaign would constrain the funnel opening angle, the duty cycle, and ultimately the EMRI+disk model.

\section{Conclusions} \label{sec:conclusions}
We investigated the early evolution of QPE associated with TDEs. We have calculated the QPE properties---such as duration, luminosity, and temperature--- within the framework of the ``EMRI+disk'' model of \citetalias{Linial&Metzger2023}. While in the original model, the accretion disk is assumed to be geometrically thin and radiatively efficient, the disk in the early stage of a TDE is likely geometrically thick and hot, so-called the slim disk. By assuming that such a slim disk is promptly formed in the TDE, we calculate the QPE observables and their time evolution. While we consider two types of disk models, the results are qualitatively similar. Our main findings are summarized as follows:

\begin{itemize}
    \item During the slim-disk phase, the disk scale height is comparable to the radius and the surface density decreases. Accordingly the QPE duration decreases over time, luminosity remains constant, and temperature increases over time. This characteristic evolution continues until the accretion rate drops below the Eddington limit. While time dependence of the duration and luminosity is moderate, the temperature evolution is relatively rapid for the insufficient thermalization case.  
    \item The QPE duration is typically shorter than those of the currently observed QPEs, $\lesssim 10^{3} \, {\rm s}$. Around the disk state transition, the duration can be as short as $\sim 100 \, {\rm s}$. The QPE temperature can be $\gtrsim 10 \, {\rm keV}$ much higher than the currently observed ones, and it would be capped by $\lesssim 50 \, {\rm keV}$ because of the pair creation. This temperature is also higher than the disk temperature, and makes the QPE distinct from the bright disk component. 
    \item Given these basic properties, we find that in principle the QPEs in the slim disk are detectable by the current X-ray detectors. However, the extremely small duty cycle of $\lesssim 10^{-2}$ makes the detection challenging. In addition, optically-thick disk winds likely block or suppress the QPEs, allowing it to be detectable only from the polar direction or around the end of the super-Eddington phase.
\end{itemize}

Detection of QPEs in the early phase of the TDE contributes to understanding the disk formation in TDEs. While our disk models are motivated by the rapid circularization scenario, where the falling back debris forms the disk instantaneously via efficient dissipation processes, a detailed picture of the disk formation process is still under debate. We propose that the early-time QPE signals plays an ideal probe of the disk. For instance, if the rapid circularization indeed occurs in TDEs, we expect the QPEs are detectable even during the TDEs unless the X-ray emission is blocked by the disk wind. In the other extreme limit, when the disk formation delays due to the inefficient dissipation, the QPEs do not appear for a long time, $\gtrsim 100 \, {\rm days}$ \citep{Piran+2015,Metzger2022,Ryu+2023b,Steinberg&Stone2024,Price+2024}. We strongly encourage prompt X-ray follow-ups of optical TDEs with high-cadence and long-time baseline to catch the early QPE signals.

\section*{acknowledgements}
We thank Takashi Hosokawa for his continuous encouragement to complete this work, and Kimitake Hayasaki for helpful suggestions. We also thank Katsuaki Asano, Kunihito Ioka, Christopher Irwin, Toshihiro Kawaguchi, Tomohisa Kawashima, Shigeo S. Kimura, Itai Linial, Brian Metzger, Andrew Mummery, Yutaka Ohira, Ken Ohsuga, Hidetoshi Omiya, Jiro Shimoda, Hiroki Takeda, Indrek Vurm, and Shogo Yoshioka for fruitful discussions and comments. T.S. is supported by JST SPRING, grant No. JPMJSP2110. T.M. acknowledges supports of the Hakubi project at Kyoto University and JSPS KAKENHI (grant No. 24K17088). This research was supported by a grant from the Kyoto University Foundation. 

\section*{data availability}
The data underlying this article will be shared on reasonable request to the corresponding author.

%%%%%%%%%%%%%%%%%%%%%%%%%%%%%%%%%%%%%%%%%%%%%%%%%%
%\section*{Data Availability}

%%%%%%%%%%%%%%%%%%%% REFERENCES %%%%%%%%%%%%%%%%%%

% The best way to enter references is to use BibTeX:

\bibliographystyle{mnras}
%\bibliography{example} % if your bibtex file is called example.bib
%\bibliography{reference_suzuguchi,reference_matsumoto}
\bibliography{ref_NT}

@ARTICLE{Abramowicz+1988,
       author = {{Abramowicz}, M.~A. and {Czerny}, B. and {Lasota}, J.~P. and {Szuszkiewicz}, E.},
        title = "{Slim Accretion Disks}",
      journal = {\apj},
         year = 1988,
        month = sep,
       volume = {332},
        pages = {646},
          doi = {10.1086/166683},
       adsurl = {https://ui.adsabs.harvard.edu/abs/1988ApJ...332..646A}
}

@ARTICLE{Alexander+2003,
       author = {{Alexander}, D.~M. and {Bauer}, F.~E. and {Brandt}, W.~N. and {Schneider}, D.~P. and {Hornschemeier}, A.~E. and {Vignali}, C. and {Barger}, A.~J. and {Broos}, P.~S. and {Cowie}, L.~L. and {Garmire}, G.~P. and {Townsley}, L.~K. and {Bautz}, M.~W. and {Chartas}, G. and {Sargent}, W.~L.~W.},
        title = "{The Chandra Deep Field North Survey. XIII. 2 Ms Point-Source Catalogs}",
      journal = {\aj},
         year = 2003,
        month = aug,
       volume = {126},
       number = {2},
        pages = {539-574},
          doi = {10.1086/376473},
archivePrefix = {arXiv},
       eprint = {astro-ph/0304392},
 primaryClass = {astro-ph},
       adsurl = {https://ui.adsabs.harvard.edu/abs/2003AJ....126..539A}
}

@ARTICLE{Alush&Stone2025,
       author = {{Alush}, Yael and {Stone}, Nicholas C.},
        title = "{Late-Time Evolution of Magnetized Disks in Tidal Disruption Events}",
      journal = {arXiv e-prints},
         year = 2025,
        month = mar,
          eid = {arXiv:2503.03811},
        pages = {arXiv:2503.03811},
          doi = {10.48550/arXiv.2503.03811},
archivePrefix = {arXiv},
       eprint = {2503.03811},
 primaryClass = {astro-ph.HE},
       adsurl = {https://ui.adsabs.harvard.edu/abs/2025arXiv250303811A}
}

@ARTICLE{Arcodia+2021,
       author = {{Arcodia}, R. and {Merloni}, A. and {Nandra}, K. and {Buchner}, J. and {Salvato}, M. and {Pasham}, D. and {Remillard}, R. and {Comparat}, J. and {Lamer}, G. and {Ponti}, G. and {Malyali}, A. and {Wolf}, J. and {Arzoumanian}, Z. and {Bogensberger}, D. and {Buckley}, D.~A.~H. and {Gendreau}, K. and {Gromadzki}, M. and {Kara}, E. and {Krumpe}, M. and {Markwardt}, C. and {Ramos-Ceja}, M.~E. and {Rau}, A. and {Schramm}, M. and {Schwope}, A.},
        title = "{X-ray quasi-periodic eruptions from two previously quiescent galaxies}",
      journal = {\nat},
         year = 2021,
        month = apr,
       volume = {592},
       number = {7856},
        pages = {704-707},
          doi = {10.1038/s41586-021-03394-6},
archivePrefix = {arXiv},
       eprint = {2104.13388},
 primaryClass = {astro-ph.HE},
       adsurl = {https://ui.adsabs.harvard.edu/abs/2021Natur.592..704A}
}

@ARTICLE{Arcodia+2024b,
       author = {{Arcodia}, R. and {Merloni}, A. and {Buchner}, J. and {Baldini}, P. and {Ponti}, G. and {Rau}, A. and {Liu}, Z. and {Nandra}, K. and {Salvato}, M.},
        title = "{Cosmic hide and seek: The volumetric rate of X-ray quasi-periodic eruptions}",
      journal = {\aap},
         year = 2024,
        month = apr,
       volume = {684},
          eid = {L14},
        pages = {L14},
          doi = {10.1051/0004-6361/202348949},
archivePrefix = {arXiv},
       eprint = {2403.17059},
 primaryClass = {astro-ph.HE},
       adsurl = {https://ui.adsabs.harvard.edu/abs/2024A&A...684L..14A}
}

@ARTICLE{Arcodia+2025,
       author = {{Arcodia}, R. and {Baldini}, P. and {Merloni}, A. and {Rau}, A. and {Nandra}, K. and {Chakraborty}, J. and {Goodwin}, A.~J. and {Page}, M.~J. and {Buchner}, J. and {Masterson}, M. and {Monageng}, I. and {Arzoumanian}, Z. and {Buckley}, D. and {Kara}, E. and {Ponti}, G. and {Ramos-Ceja}, M.~E. and {Salvato}, M. and {Gendreau}, K. and {Grotova}, I. and {Krumpe}, M.},
        title = "{SRG/eROSITA No. 5: Discovery of Quasiperiodic Eruptions Every {\ensuremath{\sim}}3.7 days from a Galaxy at z > 0.1}",
      journal = {\apj},
         year = 2025,
        month = aug,
       volume = {989},
       number = {1},
          eid = {13},
        pages = {13},
          doi = {10.3847/1538-4357/adec9b},
archivePrefix = {arXiv},
       eprint = {2506.17138},
 primaryClass = {astro-ph.HE},
       adsurl = {https://ui.adsabs.harvard.edu/abs/2025ApJ...989...13A}
}

@ARTICLE{Arnett1980,
       author = {{Arnett}, W.~D.},
        title = "{Analytic solutions for light curves of supernovae of Type II}",
      journal = {\apj},
         year = 1980,
        month = apr,
       volume = {237},
        pages = {541-549},
          doi = {10.1086/157898},
       adsurl = {https://ui.adsabs.harvard.edu/abs/1980ApJ...237..541A}
}

@ARTICLE{Begelman1978,
       author = {{Begelman}, M.~C.},
        title = "{Black holes in radiation-dominated gas: an analogue of the Bondi accretion problem.}",
      journal = {\mnras},
         year = 1978,
        month = jul,
       volume = {184},
        pages = {53-67},
          doi = {10.1093/mnras/184.1.53},
       adsurl = {https://ui.adsabs.harvard.edu/abs/1978MNRAS.184...53B}
}

@ARTICLE{Begelman&Meier1982,
       author = {{Begelman}, M.~C. and {Meier}, D.~L.},
        title = "{Thick accretion disks - Self-similar, supercritical models}",
      journal = {\apj},
         year = 1982,
        month = feb,
       volume = {253},
        pages = {873-896},
          doi = {10.1086/159688},
       adsurl = {https://ui.adsabs.harvard.edu/abs/1982ApJ...253..873B}
}

@ARTICLE{Beloborodov1998,
       author = {{Beloborodov}, A.~M.},
        title = "{Super-Eddington accretion discs around Kerr black holes}",
      journal = {\mnras},
         year = 1998,
        month = jul,
       volume = {297},
       number = {3},
        pages = {739-746},
          doi = {10.1046/j.1365-8711.1998.01530.x},
archivePrefix = {arXiv},
       eprint = {astro-ph/9802129},
 primaryClass = {astro-ph},
       adsurl = {https://ui.adsabs.harvard.edu/abs/1998MNRAS.297..739B}
}

@ARTICLE{Blandford&Begelman1999,
       author = {{Blandford}, Roger D. and {Begelman}, Mitchell C.},
        title = "{On the fate of gas accreting at a low rate on to a black hole}",
      journal = {\mnras},
         year = 1999,
        month = feb,
       volume = {303},
       number = {1},
        pages = {L1-L5},
          doi = {10.1046/j.1365-8711.1999.02358.x},
archivePrefix = {arXiv},
       eprint = {astro-ph/9809083},
 primaryClass = {astro-ph},
       adsurl = {https://ui.adsabs.harvard.edu/abs/1999MNRAS.303L...1B}
}

@ARTICLE{Bonnerot+2016,
       author = {{Bonnerot}, Cl{\'e}ment and {Rossi}, Elena M. and {Lodato}, Giuseppe and {Price}, Daniel J.},
        title = "{Disc formation from tidal disruptions of stars on eccentric orbits by Schwarzschild black holes}",
      journal = {\mnras},
         year = 2016,
        month = jan,
       volume = {455},
       number = {2},
        pages = {2253-2266},
          doi = {10.1093/mnras/stv2411},
archivePrefix = {arXiv},
       eprint = {1501.04635},
 primaryClass = {astro-ph.HE},
       adsurl = {https://ui.adsabs.harvard.edu/abs/2016MNRAS.455.2253B}
}

@ARTICLE{Bonnerot&Stone2021,
       author = {{Bonnerot}, C. and {Stone}, N.~C.},
        title = "{Formation of an Accretion Flow}",
      journal = {\ssr},
         year = 2021,
        month = feb,
       volume = {217},
       number = {1},
          eid = {16},
        pages = {16},
          doi = {10.1007/s11214-020-00789-1},
archivePrefix = {arXiv},
       eprint = {2008.11731},
 primaryClass = {astro-ph.HE},
       adsurl = {https://ui.adsabs.harvard.edu/abs/2021SSRv..217...16B}
}

@ARTICLE{Budnik+2010,
       author = {{Budnik}, Ran and {Katz}, Boaz and {Sagiv}, Amir and {Waxman}, Eli},
        title = "{Relativistic Radiation Mediated Shocks}",
      journal = {\apj},
         year = 2010,
        month = dec,
       volume = {725},
       number = {1},
        pages = {63-90},
          doi = {10.1088/0004-637X/725/1/63},
archivePrefix = {arXiv},
       eprint = {1005.0141},
 primaryClass = {astro-ph.HE},
       adsurl = {https://ui.adsabs.harvard.edu/abs/2010ApJ...725...63B}
}

@ARTICLE{Burrows+2005,
       author = {{Burrows}, David N. and {Hill}, J.~E. and {Nousek}, J.~A. and {Kennea}, J.~A. and {Wells}, A. and {Osborne}, J.~P. and {Abbey}, A.~F. and {Beardmore}, A. and {Mukerjee}, K. and {Short}, A.~D.~T. and {Chincarini}, G. and {Campana}, S. and {Citterio}, O. and {Moretti}, A. and {Pagani}, C. and {Tagliaferri}, G. and {Giommi}, P. and {Capalbi}, M. and {Tamburelli}, F. and {Angelini}, L. and {Cusumano}, G. and {Br{\"a}uninger}, H.~W. and {Burkert}, W. and {Hartner}, G.~D.},
        title = "{The Swift X-Ray Telescope}",
      journal = {\ssr},
         year = 2005,
        month = oct,
       volume = {120},
       number = {3-4},
        pages = {165-195},
          doi = {10.1007/s11214-005-5097-2},
archivePrefix = {arXiv},
       eprint = {astro-ph/0508071},
 primaryClass = {astro-ph},
       adsurl = {https://ui.adsabs.harvard.edu/abs/2005SSRv..120..165B}
}

@ARTICLE{Bykov+2025,
       author = {{Bykov}, S.~D. and {Gilfanov}, M.~R. and {Sunyaev}, R.~A. and {Medvedev}, P.~S.},
        title = "{Further evidence of quasi-periodic eruptions in a tidal disruption event AT2019vcb by SRG/eROSITA}",
      journal = {\mnras},
         year = 2025,
        month = jun,
       volume = {540},
       number = {1},
        pages = {30-36},
          doi = {10.1093/mnras/staf686},
archivePrefix = {arXiv},
       eprint = {2409.16908},
 primaryClass = {astro-ph.HE},
       adsurl = {https://ui.adsabs.harvard.edu/abs/2025MNRAS.540...30B}
}

@ARTICLE{Cannizzo+1990,
       author = {{Cannizzo}, John K. and {Lee}, Hyung Mok and {Goodman}, Jeremy},
        title = "{The Disk Accretion of a Tidally Disrupted Star onto a Massive Black Hole}",
      journal = {\apj},
         year = 1990,
        month = mar,
       volume = {351},
        pages = {38},
          doi = {10.1086/168442},
       adsurl = {https://ui.adsabs.harvard.edu/abs/1990ApJ...351...38C}
}

@ARTICLE{Cannizzo&Gehrels2009,
       author = {{Cannizzo}, J.~K. and {Gehrels}, N.},
        title = "{A New Paradigm for Gamma-ray Bursts: Long-term Accretion Rate Modulation by an External Accretion Disk}",
      journal = {\apj},
         year = 2009,
        month = aug,
       volume = {700},
       number = {2},
        pages = {1047-1058},
          doi = {10.1088/0004-637X/700/2/1047},
archivePrefix = {arXiv},
       eprint = {0901.3564},
 primaryClass = {astro-ph.HE},
       adsurl = {https://ui.adsabs.harvard.edu/abs/2009ApJ...700.1047C}
}

@ARTICLE{Cappelluti+2024,
       author = {{Cappelluti}, Nico and {Foord}, Adi and {Marchesi}, Stefano and {Pacucci}, Fabio and {Ricarte}, Angelo and {Habouzit}, Melanie and {Vito}, Fabio and {Powell}, Meredith and {Koss}, Michael and {Mushotzky}, Richard},
        title = "{Surveying the Onset and Evolution of Supermassive Black Holes at High-z with AXIS}",
      journal = {Universe},
         year = 2024,
        month = jun,
       volume = {10},
       number = {7},
          eid = {276},
        pages = {276},
          doi = {10.3390/universe10070276},
archivePrefix = {arXiv},
       eprint = {2311.07669},
 primaryClass = {astro-ph.HE},
       adsurl = {https://ui.adsabs.harvard.edu/abs/2024Univ...10..276C}
}

@ARTICLE{Chakraborty+2021,
       author = {{Chakraborty}, Joheen and {Kara}, Erin and {Masterson}, Megan and {Giustini}, Margherita and {Miniutti}, Giovanni and {Saxton}, Richard},
        title = "{Possible X-Ray Quasi-periodic Eruptions in a Tidal Disruption Event Candidate}",
      journal = {\apjl},
         year = 2021,
        month = nov,
       volume = {921},
       number = {2},
          eid = {L40},
        pages = {L40},
          doi = {10.3847/2041-8213/ac313b},
archivePrefix = {arXiv},
       eprint = {2110.10786},
 primaryClass = {astro-ph.HE},
       adsurl = {https://ui.adsabs.harvard.edu/abs/2021ApJ...921L..40C}
}

@ARTICLE{Chakraborty+2025a,
       author = {{Chakraborty}, Joheen and {Kara}, Erin and {Arcodia}, Riccardo and {Buchner}, Johannes and {Giustini}, Margherita and {Hern\textbackslash'andez-Garc\textbackslash'ia}, Lorena and {Linial}, Itai and {Masterson}, Megan and {Miniutti}, Giovanni and {Mummery}, Andrew and {Panagiotou}, Christos and {Quintin}, Erwan and {S\textbackslash'anchez-S\textbackslash'aez}, Paula},
        title = "{Discovery of Quasi-periodic Eruptions in the Tidal Disruption Event and Extreme Coronal Line Emitter AT2022upj: implications for the QPE/TDE fraction and a connection to ECLEs}",
      journal = {arXiv e-prints},
         year = 2025,
        month = mar,
          eid = {arXiv:2503.19013},
        pages = {arXiv:2503.19013},
          doi = {10.48550/arXiv.2503.19013},
archivePrefix = {arXiv},
       eprint = {2503.19013},
 primaryClass = {astro-ph.HE},
       adsurl = {https://ui.adsabs.harvard.edu/abs/2025arXiv250319013C}
}

@ARTICLE{Chakraborty+2025b,
       author = {{Chakraborty}, Joheen and {Kosec}, Peter and {Kara}, Erin and {Miniutti}, Giovanni and {Arcodia}, Riccardo and {Behar}, Ehud and {Giustini}, Margherita and {Hern{\'a}ndez-Garc{\'\i}a}, Lorena and {Masterson}, Megan and {Quintin}, Erwan and {Ricci}, Claudio and {S{\'a}nchez-S{\'a}ez}, Paula},
        title = "{Rapidly Varying Ionization Features in a Quasi-periodic Eruption: A Homologous Expansion Model for the Spectroscopic Evolution}",
      journal = {\apj},
         year = 2025,
        month = may,
       volume = {984},
       number = {2},
          eid = {124},
        pages = {124},
          doi = {10.3847/1538-4357/adb972},
archivePrefix = {arXiv},
       eprint = {2504.07167},
 primaryClass = {astro-ph.HE},
       adsurl = {https://ui.adsabs.harvard.edu/abs/2025ApJ...984..124C}
}

@ARTICLE{Chen+2022,
       author = {{Chen}, Xian and {Qiu}, Yu and {Li}, Shuo and {Liu}, F.~K.},
        title = "{Milli-Hertz Gravitational-wave Background Produced by Quasiperiodic Eruptions}",
      journal = {\apj},
         year = 2022,
        month = may,
       volume = {930},
       number = {2},
          eid = {122},
        pages = {122},
          doi = {10.3847/1538-4357/ac63bf},
archivePrefix = {arXiv},
       eprint = {2112.03408},
 primaryClass = {astro-ph.HE},
       adsurl = {https://ui.adsabs.harvard.edu/abs/2022ApJ...930..122C}
}

@ARTICLE{Dai+2010,
       author = {{Dai}, Lixin Jane and {Fuerst}, Steven V. and {Blandford}, Roger},
        title = "{Quasi-periodic flares from star-accretion-disc collisions}",
      journal = {\mnras},
         year = 2010,
        month = mar,
       volume = {402},
       number = {3},
        pages = {1614-1624},
          doi = {10.1111/j.1365-2966.2009.16038.x},
archivePrefix = {arXiv},
       eprint = {0906.0800},
 primaryClass = {astro-ph.HE},
       adsurl = {https://ui.adsabs.harvard.edu/abs/2010MNRAS.402.1614D}
}

@ARTICLE{Dai+2018,
       author = {{Dai}, Lixin and {McKinney}, Jonathan C. and {Roth}, Nathaniel and {Ramirez-Ruiz}, Enrico and {Miller}, M. Coleman},
        title = "{A Unified Model for Tidal Disruption Events}",
      journal = {\apjl},
         year = 2018,
        month = jun,
       volume = {859},
       number = {2},
          eid = {L20},
        pages = {L20},
          doi = {10.3847/2041-8213/aab429},
archivePrefix = {arXiv},
       eprint = {1803.03265},
 primaryClass = {astro-ph.HE},
       adsurl = {https://ui.adsabs.harvard.edu/abs/2018ApJ...859L..20D}
}

@ARTICLE{Duque+2025,
       author = {{Duque}, Francisco and {Kejriwal}, Shubham and {Sberna}, Laura and {Speri}, Lorenzo and {Gair}, Jonathan},
        title = "{Constraining accretion physics with gravitational waves from eccentric extreme-mass-ratio inspirals}",
      journal = {\prd},
         year = 2025,
        month = apr,
       volume = {111},
       number = {8},
          eid = {084006},
        pages = {084006},
          doi = {10.1103/PhysRevD.111.084006},
archivePrefix = {arXiv},
       eprint = {2411.03436},
 primaryClass = {gr-qc},
       adsurl = {https://ui.adsabs.harvard.edu/abs/2025PhRvD.111h4006D}
}

@ARTICLE{Faran&Sari2019,
       author = {{Faran}, Tamar and {Sari}, Re'em},
        title = "{Early Supernova Emission: Logarithmic Corrections to the Planar Phase}",
      journal = {\apj},
         year = 2019,
        month = oct,
       volume = {884},
       number = {1},
          eid = {41},
        pages = {41},
          doi = {10.3847/1538-4357/ab3e3d},
archivePrefix = {arXiv},
       eprint = {1908.06990},
 primaryClass = {astro-ph.HE},
       adsurl = {https://ui.adsabs.harvard.edu/abs/2019ApJ...884...41F}
}

@ARTICLE{Franchini+2023,
       author = {{Franchini}, Alessia and {Bonetti}, Matteo and {Lupi}, Alessandro and {Miniutti}, Giovanni and {Bortolas}, Elisa and {Giustini}, Margherita and {Dotti}, Massimo and {Sesana}, Alberto and {Arcodia}, Riccardo and {Ryu}, Taeho},
        title = "{Quasi-periodic eruptions from impacts between the secondary and a rigidly precessing accretion disc in an extreme mass-ratio inspiral system}",
      journal = {\aap},
         year = 2023,
        month = jul,
       volume = {675},
          eid = {A100},
        pages = {A100},
          doi = {10.1051/0004-6361/202346565},
archivePrefix = {arXiv},
       eprint = {2304.00775},
 primaryClass = {astro-ph.HE},
       adsurl = {https://ui.adsabs.harvard.edu/abs/2023A&A...675A.100F}
}

@ARTICLE{Generozov&Perets2023,
       author = {{Generozov}, A. and {Perets}, H.~B.},
        title = "{Capture of stars into gaseous discs around massive black holes: alignment, circularization, and growth}",
      journal = {\mnras},
         year = 2023,
        month = jun,
       volume = {522},
       number = {2},
        pages = {1763-1778},
          doi = {10.1093/mnras/stad1016},
archivePrefix = {arXiv},
       eprint = {2212.11301},
 primaryClass = {astro-ph.GA},
       adsurl = {https://ui.adsabs.harvard.edu/abs/2023MNRAS.522.1763G}
}

@ARTICLE{Gezari+2017,
       author = {{Gezari}, S. and {Cenko}, S.~B. and {Arcavi}, I.},
        title = "{X-Ray Brightening and UV Fading of Tidal Disruption Event ASASSN-15oi}",
      journal = {\apjl},
         year = 2017,
        month = dec,
       volume = {851},
       number = {2},
          eid = {L47},
        pages = {L47},
          doi = {10.3847/2041-8213/aaa0c2},
archivePrefix = {arXiv},
       eprint = {1712.03968},
 primaryClass = {astro-ph.HE},
       adsurl = {https://ui.adsabs.harvard.edu/abs/2017ApJ...851L..47G}
}

@ARTICLE{Gezari2021,
       author = {{Gezari}, Suvi},
        title = "{Tidal Disruption Events}",
      journal = {\araa},
         year = 2021,
        month = sep,
       volume = {59},
        pages = {21-58},
          doi = {10.1146/annurev-astro-111720-030029},
archivePrefix = {arXiv},
       eprint = {2104.14580},
 primaryClass = {astro-ph.HE},
       adsurl = {https://ui.adsabs.harvard.edu/abs/2021ARA&A..59...21G}
}

@ARTICLE{Giustini+2020,
       author = {{Giustini}, Margherita and {Miniutti}, Giovanni and {Saxton}, Richard D.},
        title = "{X-ray quasi-periodic eruptions from the galactic nucleus of RX J1301.9+2747}",
      journal = {\aap},
         year = 2020,
        month = apr,
       volume = {636},
          eid = {L2},
        pages = {L2},
          doi = {10.1051/0004-6361/202037610},
archivePrefix = {arXiv},
       eprint = {2002.08967},
 primaryClass = {astro-ph.HE},
       adsurl = {https://ui.adsabs.harvard.edu/abs/2020A&A...636L...2G}
}

@ARTICLE{Grotova+2025,
       author = {{Grotova}, I. and {Rau}, A. and {Baldini}, P. and {Goodwin}, A.~J. and {Liu}, Z. and {Merloni}, A. and {Salvato}, M. and {Anderson}, G.~E. and {Arcodia}, R. and {Buchner}, J. and {Krumpe}, M. and {Malyali}, A. and {Masterson}, M. and {Miller-Jones}, J.~C.~A. and {Nandra}, K. and {Shirley}, R.},
        title = "{The population of tidal disruption events discovered with eROSITA}",
      journal = {\aap},
         year = 2025,
        month = may,
       volume = {697},
          eid = {A159},
        pages = {A159},
          doi = {10.1051/0004-6361/202553669},
archivePrefix = {arXiv},
       eprint = {2504.08424},
 primaryClass = {astro-ph.HE},
       adsurl = {https://ui.adsabs.harvard.edu/abs/2025A&A...697A.159G}
}

@ARTICLE{Guo&Shen2025,
       author = {{Guo}, Wenyuan and {Shen}, Rong-Feng},
        title = "{Testing the Star-disk Collision Model for Quasi-periodic Eruptions}",
      journal = {arXiv e-prints},
         year = 2025,
        month = apr,
          eid = {arXiv:2504.12762},
        pages = {arXiv:2504.12762},
          doi = {10.48550/arXiv.2504.12762},
archivePrefix = {arXiv},
       eprint = {2504.12762},
 primaryClass = {astro-ph.HE},
       adsurl = {https://ui.adsabs.harvard.edu/abs/2025arXiv250412762G}
}

@ARTICLE{Guolo+2024,
       author = {{Guolo}, Muryel and {Gezari}, Suvi and {Yao}, Yuhan and {van Velzen}, Sjoert and {Hammerstein}, Erica and {Cenko}, S. Bradley and {Tokayer}, Yarone M.},
        title = "{A Systematic Analysis of the X-Ray Emission in Optically Selected Tidal Disruption Events: Observational Evidence for the Unification of the Optically and X-Ray-selected Populations}",
      journal = {\apj},
         year = 2024,
        month = may,
       volume = {966},
       number = {2},
          eid = {160},
        pages = {160},
          doi = {10.3847/1538-4357/ad2f9f},
archivePrefix = {arXiv},
       eprint = {2308.13019},
 primaryClass = {astro-ph.HE},
       adsurl = {https://ui.adsabs.harvard.edu/abs/2024ApJ...966..160G}
}

@ARTICLE{Guolo+2025,
       author = {{Guolo}, M. and {Mummery}, A. and {Wevers}, T. and {Nicholl}, M. and {Gezari}, S. and {Ingram}, A. and {Pasham}, D.~R.},
        title = "{Properties of the GSN 069 Accretion Disk from a Joint X-Ray and UV Spectral Analysis: Stress-testing Quasi-periodic Eruption Models}",
      journal = {\apj},
         year = 2025,
        month = jun,
       volume = {985},
       number = {2},
          eid = {146},
        pages = {146},
          doi = {10.3847/1538-4357/adcbac},
archivePrefix = {arXiv},
       eprint = {2501.03333},
 primaryClass = {astro-ph.HE},
       adsurl = {https://ui.adsabs.harvard.edu/abs/2025ApJ...985..146G}
}

@ARTICLE{Hasinger+2001,
       author = {{Hasinger}, G. and {Altieri}, B. and {Arnaud}, M. and {Barcons}, X. and {Bergeron}, J. and {Brunner}, H. and {Dadina}, M. and {Dennerl}, K. and {Ferrando}, P. and {Finoguenov}, A. and {Griffiths}, R.~E. and {Hashimoto}, Y. and {Jansen}, F.~A. and {Lumb}, D.~H. and {Mason}, K.~O. and {Mateos}, S. and {McMahon}, R.~G. and {Miyaji}, T. and {Paerels}, F. and {Page}, M.~J. and {Ptak}, A.~F. and {Sasseen}, T.~P. and {Schartel}, N. and {Szokoly}, G.~P. and {Tr{\"u}mper}, J. and {Turner}, M. and {Warwick}, R.~S. and {Watson}, M.~G.},
        title = "{XMM-Newton observation of the Lockman Hole. I. The X-ray data}",
      journal = {\aap},
         year = 2001,
        month = jan,
       volume = {365},
        pages = {L45-L50},
          doi = {10.1051/0004-6361:20000046},
archivePrefix = {arXiv},
       eprint = {astro-ph/0011271},
 primaryClass = {astro-ph},
       adsurl = {https://ui.adsabs.harvard.edu/abs/2001A&A...365L..45H}
}

@ARTICLE{Hayasaki+2013,
       author = {{Hayasaki}, Kimitake and {Stone}, Nicholas and {Loeb}, Abraham},
        title = "{Finite, intense accretion bursts from tidal disruption of stars on bound orbits}",
      journal = {\mnras},
         year = 2013,
        month = sep,
       volume = {434},
       number = {2},
        pages = {909-924},
          doi = {10.1093/mnras/stt871},
archivePrefix = {arXiv},
       eprint = {1210.1333},
 primaryClass = {astro-ph.HE},
       adsurl = {https://ui.adsabs.harvard.edu/abs/2013MNRAS.434..909H}
}

@ARTICLE{Hayasaki+2016,
       author = {{Hayasaki}, Kimitake and {Stone}, Nicholas and {Loeb}, Abraham},
        title = "{Circularization of tidally disrupted stars around spinning supermassive black holes}",
      journal = {\mnras},
         year = 2016,
        month = oct,
       volume = {461},
       number = {4},
        pages = {3760-3780},
          doi = {10.1093/mnras/stw1387},
archivePrefix = {arXiv},
       eprint = {1501.05207},
 primaryClass = {astro-ph.HE},
       adsurl = {https://ui.adsabs.harvard.edu/abs/2016MNRAS.461.3760H}
}

@ARTICLE{Hernandez-Garcia+2025,
       author = {{Hern{\'a}ndez-Garc{\'\i}a}, Lorena and {Chakraborty}, Joheen and {S{\'a}nchez-S{\'a}ez}, Paula and {Ricci}, Claudio and {Cuadra}, Jorge and {McKernan}, Barry and {Ford}, K.~E. Saavik and {Ar{\'e}valo}, Patricia and {Rau}, Arne and {Arcodia}, Riccardo and {Kara}, Erin and {Liu}, Zhu and {Merloni}, Andrea and {Bruni}, Gabriele and {Goodwin}, Adelle and {Arzoumanian}, Zaven and {Assef}, Roberto J. and {Baldini}, Pietro and {Bayo}, Amelia and {Bauer}, Franz E. and {Bernal}, Santiago and {Brightman}, Murray and {Calistro Rivera}, Gabriela and {Gendreau}, Keith and {Homan}, David and {Krumpe}, Mirko and {Lira}, Paulina and {Mart{\'\i}nez-Aldama}, Mary Loli and {Salvato}, Mara and {Sotomayor}, Bel{\'e}n},
        title = "{Discovery of extreme Quasi-Periodic Eruptions in a newly accreting massive black hole}",
      journal = {arXiv e-prints},
         year = 2025,
        month = apr,
          eid = {arXiv:2504.07169},
        pages = {arXiv:2504.07169},
          doi = {10.48550/arXiv.2504.07169},
archivePrefix = {arXiv},
       eprint = {2504.07169},
 primaryClass = {astro-ph.HE},
       adsurl = {https://ui.adsabs.harvard.edu/abs/2025arXiv250407169H}
}

@ARTICLE{Hills1975,
       author = {{Hills}, J.~G.},
        title = "{Possible power source of Seyfert galaxies and QSOs}",
      journal = {\nat},
         year = 1975,
        month = mar,
       volume = {254},
       number = {5498},
        pages = {295-298},
          doi = {10.1038/254295a0},
       adsurl = {https://ui.adsabs.harvard.edu/abs/1975Natur.254..295H}
}

@ARTICLE{Holoien+2016,
       author = {{Holoien}, T.~W. -S. and {Kochanek}, C.~S. and {Prieto}, J.~L. and {Grupe}, D. and {Chen}, Ping and {Godoy-Rivera}, D. and {Stanek}, K.~Z. and {Shappee}, B.~J. and {Dong}, Subo and {Brown}, J.~S. and {Basu}, U. and {Beacom}, J.~F. and {Bersier}, D. and {Brimacombe}, J. and {Carlson}, E.~K. and {Falco}, E. and {Johnston}, E. and {Madore}, B.~F. and {Pojmanski}, G. and {Seibert}, M.},
        title = "{ASASSN-15oi: a rapidly evolving, luminous tidal disruption event at 216 Mpc}",
      journal = {\mnras},
         year = 2016,
        month = dec,
       volume = {463},
       number = {4},
        pages = {3813-3828},
          doi = {10.1093/mnras/stw2272},
archivePrefix = {arXiv},
       eprint = {1602.01088},
 primaryClass = {astro-ph.HE},
       adsurl = {https://ui.adsabs.harvard.edu/abs/2016MNRAS.463.3813H}
}

@ARTICLE{Huang+2025,
       author = {{Huang}, Xiaoshan and {Linial}, Itai and {Jiang}, Yan-Fei},
        title = "{Multi-band Emission from Star-Disk Collision and Implications for Quasi-Periodic Eruptions}",
      journal = {arXiv e-prints},
         year = 2025,
        month = jun,
          eid = {arXiv:2506.11231},
        pages = {arXiv:2506.11231},
          doi = {10.48550/arXiv.2506.11231},
archivePrefix = {arXiv},
       eprint = {2506.11231},
 primaryClass = {astro-ph.HE},
       adsurl = {https://ui.adsabs.harvard.edu/abs/2025arXiv250611231H}
}

@ARTICLE{Ingram+2021,
       author = {{Ingram}, Adam and {Motta}, Sara E. and {Aigrain}, Suzanne and {Karastergiou}, Aris},
        title = "{A self-lensing binary massive black hole interpretation of quasi-periodic eruptions}",
      journal = {\mnras},
         year = 2021,
        month = may,
       volume = {503},
       number = {2},
        pages = {1703-1716},
          doi = {10.1093/mnras/stab609},
archivePrefix = {arXiv},
       eprint = {2103.00017},
 primaryClass = {astro-ph.HE},
       adsurl = {https://ui.adsabs.harvard.edu/abs/2021MNRAS.503.1703I}
}

@ARTICLE{Irwin&Hotokezaka2024,
       author = {{Irwin}, Christopher M. and {Hotokezaka}, Kenta},
        title = "{An unexplored regime of shock breakout with a distinct spectral signature}",
      journal = {arXiv e-prints},
         year = 2024,
        month = dec,
          eid = {arXiv:2412.06733},
        pages = {arXiv:2412.06733},
          doi = {10.48550/arXiv.2412.06733},
archivePrefix = {arXiv},
       eprint = {2412.06733},
 primaryClass = {astro-ph.HE},
       adsurl = {https://ui.adsabs.harvard.edu/abs/2024arXiv241206733I}
}

@ARTICLE{Ivanov+1998,
       author = {{Ivanov}, Pavel B. and {Igumenshchev}, Igor V. and {Novikov}, Igor D.},
        title = "{Hydrodynamics of Black Hole-Accretion Disk Collision}",
      journal = {\apj},
         year = 1998,
        month = nov,
       volume = {507},
       number = {1},
        pages = {131-144},
          doi = {10.1086/306324},
       adsurl = {https://ui.adsabs.harvard.edu/abs/1998ApJ...507..131I}
}

@ARTICLE{Jiang+2019,
       author = {{Jiang}, Yan-Fei and {Stone}, James M. and {Davis}, Shane W.},
        title = "{Super-Eddington Accretion Disks around Supermassive Black Holes}",
      journal = {\apj},
         year = 2019,
        month = aug,
       volume = {880},
       number = {2},
          eid = {67},
        pages = {67},
          doi = {10.3847/1538-4357/ab29ff},
archivePrefix = {arXiv},
       eprint = {1709.02845},
 primaryClass = {astro-ph.HE},
       adsurl = {https://ui.adsabs.harvard.edu/abs/2019ApJ...880...67J}
}

@BOOK{Kato+2008,
       author = {{Kato}, S. and {Fukue}, J. and {Mineshige}, S.},
        title = "{Black-Hole Accretion Disks --- Towards a New Paradigm ---}",
         year = 2008,
       adsurl = {https://ui.adsabs.harvard.edu/abs/2008bhad.book.....K}
}

@ARTICLE{Katz+2010,
       author = {{Katz}, Boaz and {Budnik}, Ran and {Waxman}, Eli},
        title = "{Fast Radiation Mediated Shocks and Supernova Shock Breakouts}",
      journal = {\apj},
         year = 2010,
        month = jun,
       volume = {716},
       number = {1},
        pages = {781-791},
          doi = {10.1088/0004-637X/716/1/781},
archivePrefix = {arXiv},
       eprint = {0902.4708},
 primaryClass = {astro-ph.HE},
       adsurl = {https://ui.adsabs.harvard.edu/abs/2010ApJ...716..781K}
}

@ARTICLE{Kaur+2023,
       author = {{Kaur}, Karamveer and {Stone}, Nicholas C. and {Gilbaum}, Shmuel},
        title = "{Magnetically dominated discs in tidal disruption events and quasi-periodic eruptions}",
      journal = {\mnras},
         year = 2023,
        month = sep,
       volume = {524},
       number = {1},
        pages = {1269-1290},
          doi = {10.1093/mnras/stad1894},
archivePrefix = {arXiv},
       eprint = {2211.00704},
 primaryClass = {astro-ph.HE},
       adsurl = {https://ui.adsabs.harvard.edu/abs/2023MNRAS.524.1269K}
}

@ARTICLE{Kejriwal+2024,
       author = {{Kejriwal}, Shubham and {Witzany}, Vojt{\v{e}}ch and {Zaja{\v{c}}ek}, Michal and {Pasham}, Dheeraj R. and {Chua}, Alvin J.~K.},
        title = "{Repeating nuclear transients as candidate electromagnetic counterparts of LISA extreme mass ratio inspirals}",
      journal = {\mnras},
         year = 2024,
        month = aug,
       volume = {532},
       number = {2},
        pages = {2143-2158},
          doi = {10.1093/mnras/stae1599},
archivePrefix = {arXiv},
       eprint = {2404.00941},
 primaryClass = {astro-ph.HE},
       adsurl = {https://ui.adsabs.harvard.edu/abs/2024MNRAS.532.2143K}
}

@ARTICLE{King2020,
       author = {{King}, Andrew},
        title = "{GSN 069 - A tidal disruption near miss}",
      journal = {\mnras},
         year = 2020,
        month = mar,
       volume = {493},
       number = {1},
        pages = {L120-L123},
          doi = {10.1093/mnrasl/slaa020},
archivePrefix = {arXiv},
       eprint = {2002.00970},
 primaryClass = {astro-ph.HE},
       adsurl = {https://ui.adsabs.harvard.edu/abs/2020MNRAS.493L.120K}
}

@ARTICLE{King2022,
       author = {{King}, Andrew},
        title = "{Quasi-periodic eruptions from galaxy nuclei}",
      journal = {\mnras},
         year = 2022,
        month = sep,
       volume = {515},
       number = {3},
        pages = {4344-4349},
          doi = {10.1093/mnras/stac1641},
archivePrefix = {arXiv},
       eprint = {2206.04698},
 primaryClass = {astro-ph.GA},
       adsurl = {https://ui.adsabs.harvard.edu/abs/2022MNRAS.515.4344K}
}

@ARTICLE{Komossa2015,
       author = {{Komossa}, S.},
        title = "{Tidal disruption of stars by supermassive black holes: Status of observations}",
      journal = {Journal of High Energy Astrophysics},
         year = 2015,
        month = sep,
       volume = {7},
        pages = {148-157},
          doi = {10.1016/j.jheap.2015.04.006},
archivePrefix = {arXiv},
       eprint = {1505.01093},
 primaryClass = {astro-ph.HE},
       adsurl = {https://ui.adsabs.harvard.edu/abs/2015JHEAp...7..148K}
}

@ARTICLE{Krolik&Linial2022,
       author = {{Krolik}, Julian H. and {Linial}, Itai},
        title = "{Quasiperiodic Erupters: A Stellar Mass-transfer Model for the Radiation}",
      journal = {\apj},
         year = 2022,
        month = dec,
       volume = {941},
       number = {1},
          eid = {24},
        pages = {24},
          doi = {10.3847/1538-4357/ac9eb6},
archivePrefix = {arXiv},
       eprint = {2209.02786},
 primaryClass = {astro-ph.HE},
       adsurl = {https://ui.adsabs.harvard.edu/abs/2022ApJ...941...24K}
}

@ARTICLE{Krolik+2025,
       author = {{Krolik}, Julian and {Piran}, Tsvi and {Ryu}, Taeho},
        title = "{Follow the Mass{\textemdash}A Concordance Picture of Tidal Disruption Events}",
      journal = {\apj},
         year = 2025,
        month = aug,
       volume = {988},
       number = {2},
          eid = {220},
        pages = {220},
          doi = {10.3847/1538-4357/ade797},
archivePrefix = {arXiv},
       eprint = {2409.02894},
 primaryClass = {astro-ph.HE},
       adsurl = {https://ui.adsabs.harvard.edu/abs/2025ApJ...988..220K}
}

@ARTICLE{Lightman&Eardley1974,
       author = {{Lightman}, Alan P. and {Eardley}, Douglas M.},
        title = "{Black Holes in Binary Systems: Instability of Disk Accretion}",
      journal = {\apjl},
         year = 1974,
        month = jan,
       volume = {187},
        pages = {L1},
          doi = {10.1086/181377},
       adsurl = {https://ui.adsabs.harvard.edu/abs/1974ApJ...187L...1L}
}

@ARTICLE{Linial&Sari2023,
       author = {{Linial}, Itai and {Sari}, Re'em},
        title = "{Unstable Mass Transfer from a Main-sequence Star to a Supermassive Black Hole and Quasiperiodic Eruptions}",
      journal = {\apj},
         year = 2023,
        month = mar,
       volume = {945},
       number = {2},
          eid = {86},
        pages = {86},
          doi = {10.3847/1538-4357/acbd3d},
archivePrefix = {arXiv},
       eprint = {2211.09851},
 primaryClass = {astro-ph.HE},
       adsurl = {https://ui.adsabs.harvard.edu/abs/2023ApJ...945...86L}
}

@ARTICLE{Linial&Metzger2023,
       author = {{Linial}, Itai and {Metzger}, Brian D.},
        title = "{EMRI + TDE = QPE: Periodic X-Ray Flares from Star-Disk Collisions in Galactic Nuclei}",
      journal = {\apj},
         year = 2023,
        month = nov,
       volume = {957},
       number = {1},
          eid = {34},
        pages = {34},
          doi = {10.3847/1538-4357/acf65b},
archivePrefix = {arXiv},
       eprint = {2303.16231},
 primaryClass = {astro-ph.HE},
       adsurl = {https://ui.adsabs.harvard.edu/abs/2023ApJ...957...34L}
}

@ARTICLE{Linial&Metzger2024b,
       author = {{Linial}, Itai and {Metzger}, Brian D.},
        title = "{Coupled Disk-star Evolution in Galactic Nuclei and the Lifetimes of QPE Sources}",
      journal = {\apj},
         year = 2024,
        month = oct,
       volume = {973},
       number = {2},
          eid = {101},
        pages = {101},
          doi = {10.3847/1538-4357/ad639e},
archivePrefix = {arXiv},
       eprint = {2404.12421},
 primaryClass = {astro-ph.HE},
       adsurl = {https://ui.adsabs.harvard.edu/abs/2024ApJ...973..101L}
}

@ARTICLE{Linial+2025,
       author = {{Linial}, Itai and {Metzger}, Brian D. and {Quataert}, Eliot},
        title = "{QPEs from EMRI Debris Streams Impacting Accretion Disks in Galactic Nuclei}",
      journal = {arXiv e-prints},
         year = 2025,
        month = jun,
          eid = {arXiv:2506.10096},
        pages = {arXiv:2506.10096},
          doi = {10.48550/arXiv.2506.10096},
archivePrefix = {arXiv},
       eprint = {2506.10096},
 primaryClass = {astro-ph.HE},
       adsurl = {https://ui.adsabs.harvard.edu/abs/2025arXiv250610096L}
}

@ARTICLE{Lops+2023,
       author = {{Lops}, Gaia and {Izquierdo-Villalba}, David and {Colpi}, Monica and {Bonoli}, Silvia and {Sesana}, Alberto and {Mangiagli}, Alberto},
        title = "{Galaxy fields of LISA massive black hole mergers in a simulated universe}",
      journal = {\mnras},
         year = 2023,
        month = mar,
       volume = {519},
       number = {4},
        pages = {5962-5986},
          doi = {10.1093/mnras/stad058},
archivePrefix = {arXiv},
       eprint = {2207.10683},
 primaryClass = {astro-ph.GA},
       adsurl = {https://ui.adsabs.harvard.edu/abs/2023MNRAS.519.5962L}
}

@ARTICLE{Lu&Bonnerot2020,
       author = {{Lu}, Wenbin and {Bonnerot}, Cl{\'e}ment},
        title = "{Self-intersection of the fallback stream in tidal disruption events}",
      journal = {\mnras},
         year = 2020,
        month = feb,
       volume = {492},
       number = {1},
        pages = {686-707},
          doi = {10.1093/mnras/stz3405},
archivePrefix = {arXiv},
       eprint = {1904.12018},
 primaryClass = {astro-ph.HE},
       adsurl = {https://ui.adsabs.harvard.edu/abs/2020MNRAS.492..686L}
}

@ARTICLE{Lu&Quataert2023,
       author = {{Lu}, Wenbin and {Quataert}, Eliot},
        title = "{Quasi-periodic eruptions from mildly eccentric unstable mass transfer in galactic nuclei}",
      journal = {\mnras},
         year = 2023,
        month = oct,
       volume = {524},
       number = {4},
        pages = {6247-6266},
          doi = {10.1093/mnras/stad2203},
archivePrefix = {arXiv},
       eprint = {2210.08023},
 primaryClass = {astro-ph.HE},
       adsurl = {https://ui.adsabs.harvard.edu/abs/2023MNRAS.524.6247L}
}

@ARTICLE{Lui+2025,
       author = {{Lui}, Leif and {Torres-Orjuela}, Alejandro and {Chowdhury}, Rudrani Kar and {Dai}, Lixin},
        title = "{Gravitational Wave Signatures of Quasi-Periodic Eruptions: LISA Detection Prospects for RX J1301.9+2747}",
      journal = {arXiv e-prints},
         year = 2025,
        month = aug,
          eid = {arXiv:2508.07961},
        pages = {arXiv:2508.07961},
          doi = {10.48550/arXiv.2508.07961},
archivePrefix = {arXiv},
       eprint = {2508.07961},
 primaryClass = {astro-ph.HE},
       adsurl = {https://ui.adsabs.harvard.edu/abs/2025arXiv250807961L}
}

@ARTICLE{Lynden-Bell&Pringle1974,
       author = {{Lynden-Bell}, D. and {Pringle}, J.~E.},
        title = "{The evolution of viscous discs and the origin of the nebular variables.}",
      journal = {\mnras},
         year = 1974,
        month = sep,
       volume = {168},
        pages = {603-637},
          doi = {10.1093/mnras/168.3.603},
       adsurl = {https://ui.adsabs.harvard.edu/abs/1974MNRAS.168..603L}
}

@ARTICLE{Lyu+2024,
       author = {{Lyu}, Zhenwei and {Pan}, Zhen and {Mao}, Junjie and {Jiang}, Ning and {Yang}, Huan},
        title = "{Science Opportunities of Wet Extreme Mass-Ratio Inspirals}",
      journal = {arXiv e-prints},
         year = 2024,
        month = dec,
          eid = {arXiv:2501.03252},
        pages = {arXiv:2501.03252},
          doi = {10.48550/arXiv.2501.03252},
archivePrefix = {arXiv},
       eprint = {2501.03252},
 primaryClass = {astro-ph.HE},
       adsurl = {https://ui.adsabs.harvard.edu/abs/2025arXiv250103252L}
}

@ARTICLE{Lyubarskij&Shakura1987,
       author = {{Lyubarskij}, Y.~E. and {Shakura}, N.~I.},
        title = "{Nonlinear self-similar problems of nonstationary disk accretion}",
      journal = {Soviet Astronomy Letters},
         year = 1987,
        month = oct,
       volume = {13},
        pages = {386},
       adsurl = {https://ui.adsabs.harvard.edu/abs/1987SvAL...13..386L}
}

@ARTICLE{Matsumoto&Piran2021,
       author = {{Matsumoto}, Tatsuya and {Piran}, Tsvi},
        title = "{Limits on mass outflow from optical tidal disruption events}",
      journal = {\mnras},
         year = 2021,
        month = apr,
       volume = {502},
       number = {3},
        pages = {3385-3393},
          doi = {10.1093/mnras/stab240},
archivePrefix = {arXiv},
       eprint = {2009.01240},
 primaryClass = {astro-ph.HE},
       adsurl = {https://ui.adsabs.harvard.edu/abs/2021MNRAS.502.3385M}
}

@ARTICLE{Metzger&Stone2016,
       author = {{Metzger}, Brian D. and {Stone}, Nicholas C.},
        title = "{A bright year for tidal disruptions}",
      journal = {\mnras},
         year = 2016,
        month = sep,
       volume = {461},
       number = {1},
        pages = {948-966},
          doi = {10.1093/mnras/stw1394},
archivePrefix = {arXiv},
       eprint = {1506.03453},
 primaryClass = {astro-ph.HE},
       adsurl = {https://ui.adsabs.harvard.edu/abs/2016MNRAS.461..948M}
}

@ARTICLE{Metzger+2022,
       author = {{Metzger}, Brian D. and {Stone}, Nicholas C. and {Gilbaum}, Shmuel},
        title = "{Interacting Stellar EMRIs as Sources of Quasi-periodic Eruptions in Galactic Nuclei}",
      journal = {\apj},
         year = 2022,
        month = feb,
       volume = {926},
       number = {1},
          eid = {101},
        pages = {101},
          doi = {10.3847/1538-4357/ac3ee1},
archivePrefix = {arXiv},
       eprint = {2107.13015},
 primaryClass = {astro-ph.HE},
       adsurl = {https://ui.adsabs.harvard.edu/abs/2022ApJ...926..101M}
}

@ARTICLE{Metzger2022,
       author = {{Metzger}, Brian D.},
        title = "{Cooling Envelope Model for Tidal Disruption Events}",
      journal = {\apjl},
         year = 2022,
        month = sep,
       volume = {937},
       number = {1},
          eid = {L12},
        pages = {L12},
          doi = {10.3847/2041-8213/ac90ba},
archivePrefix = {arXiv},
       eprint = {2207.07136},
 primaryClass = {astro-ph.HE},
       adsurl = {https://ui.adsabs.harvard.edu/abs/2022ApJ...937L..12M}
}

@ARTICLE{Middleton+2025,
       author = {{Middleton}, M. and {G{\'u}rpide}, A. and {Kwan}, T.~M. and {Dai}, L. and {Arcodia}, R. and {Chakraborty}, J. and {Dauser}, T. and {Fragile}, P.~C. and {Ingram}, A. and {Miniutti}, G. and {Pinto}, C. and {Kosec}, P.},
        title = "{Quasi-periodic eruptions as Lense-Thirring precession of super-Eddington flows}",
      journal = {\mnras},
         year = 2025,
        month = feb,
       volume = {537},
       number = {2},
        pages = {1688-1702},
          doi = {10.1093/mnras/staf052},
archivePrefix = {arXiv},
       eprint = {2501.06185},
 primaryClass = {astro-ph.HE},
       adsurl = {https://ui.adsabs.harvard.edu/abs/2025MNRAS.537.1688M}
}

@ARTICLE{Miniutti+2019,
       author = {{Miniutti}, G. and {Saxton}, R.~D. and {Giustini}, M. and {Alexander}, K.~D. and {Fender}, R.~P. and {Heywood}, I. and {Monageng}, I. and {Coriat}, M. and {Tzioumis}, A.~K. and {Read}, A.~M. and {Knigge}, C. and {Gandhi}, P. and {Pretorius}, M.~L. and {Ag{\'\i}s-Gonz{\'a}lez}, B.},
        title = "{Nine-hour X-ray quasi-periodic eruptions from a low-mass black hole galactic nucleus}",
      journal = {\nat},
         year = 2019,
        month = sep,
       volume = {573},
       number = {7774},
        pages = {381-384},
          doi = {10.1038/s41586-019-1556-x},
archivePrefix = {arXiv},
       eprint = {1909.04693},
 primaryClass = {astro-ph.HE},
       adsurl = {https://ui.adsabs.harvard.edu/abs/2019Natur.573..381M}
}

@ARTICLE{Mummery+2024,
       author = {{Mummery}, Andrew and {van Velzen}, Sjoert and {Nathan}, Edward and {Ingram}, Adam and {Hammerstein}, Erica and {Fraser-Taliente}, Ludovic and {Balbus}, Steven},
        title = "{Fundamental scaling relationships revealed in the optical light curves of tidal disruption events}",
      journal = {\mnras},
         year = 2024,
        month = jan,
       volume = {527},
       number = {2},
        pages = {2452-2489},
          doi = {10.1093/mnras/stad3001},
archivePrefix = {arXiv},
       eprint = {2308.08255},
 primaryClass = {astro-ph.HE},
       adsurl = {https://ui.adsabs.harvard.edu/abs/2024MNRAS.527.2452M}
}

@ARTICLE{Mummery2025,
       author = {{Mummery}, Andrew},
        title = "{Collisions with tidal disruption event disks: implications for quasi-periodic X-ray eruptions}",
      journal = {arXiv e-prints},
         year = 2025,
        month = apr,
          eid = {arXiv:2504.21456},
        pages = {arXiv:2504.21456},
          doi = {10.48550/arXiv.2504.21456},
archivePrefix = {arXiv},
       eprint = {2504.21456},
 primaryClass = {astro-ph.HE},
       adsurl = {https://ui.adsabs.harvard.edu/abs/2025arXiv250421456M}
}

@BOOK{Murray&Dermott1999,
       author = {{Murray}, Carl D. and {Dermott}, Stanley F.},
        title = "{Solar System Dynamics}",
         year = 1999,
          doi = {10.1017/CBO9781139174817},
       adsurl = {https://ui.adsabs.harvard.edu/abs/1999ssd..book.....M}
}

@ARTICLE{Nakar&Sari2010,
       author = {{Nakar}, Ehud and {Sari}, Re'em},
        title = "{Early Supernovae Light Curves Following the Shock Breakout}",
      journal = {\apj},
         year = 2010,
        month = dec,
       volume = {725},
       number = {1},
        pages = {904-921},
          doi = {10.1088/0004-637X/725/1/904},
archivePrefix = {arXiv},
       eprint = {1004.2496},
 primaryClass = {astro-ph.HE},
       adsurl = {https://ui.adsabs.harvard.edu/abs/2010ApJ...725..904N}
}

@ARTICLE{Nicholl+2024,
       author = {{Nicholl}, M. and {Pasham}, D.~R. and {Mummery}, A. and {Guolo}, M. and {Gendreau}, K. and {Dewangan}, G.~C. and {Ferrara}, E.~C. and {Remillard}, R. and {Bonnerot}, C. and {Chakraborty}, J. and {Hajela}, A. and {Dhillon}, V.~S. and {Gillan}, A.~F. and {Greenwood}, J. and {Huber}, M.~E. and {Janiuk}, A. and {Salvesen}, G. and {van Velzen}, S. and {Aamer}, A. and {Alexander}, K.~D. and {Angus}, C.~R. and {Arzoumanian}, Z. and {Auchettl}, K. and {Berger}, E. and {de Boer}, T. and {Cendes}, Y. and {Chambers}, K.~C. and {Chen}, T. -W. and {Chornock}, R. and {Fulton}, M.~D. and {Gao}, H. and {Gillanders}, J.~H. and {Gomez}, S. and {Gompertz}, B.~P. and {Fabian}, A.~C. and {Herman}, J. and {Ingram}, A. and {Kara}, E. and {Laskar}, T. and {Lawrence}, A. and {Lin}, C. -C. and {Lowe}, T.~B. and {Magnier}, E.~A. and {Margutti}, R. and {McGee}, S.~L. and {Minguez}, P. and {Moore}, T. and {Nathan}, E. and {Oates}, S.~R. and {Patra}, K.~C. and {Ramsden}, P. and {Ravi}, V. and {Ridley}, E.~J. and {Sheng}, X. and {Smartt}, S.~J. and {Smith}, K.~W. and {Srivastav}, S. and {Stein}, R. and {Stevance}, H.~F. and {Turner}, S.~G.~D. and {Wainscoat}, R.~J. and {Weston}, J. and {Wevers}, T. and {Young}, D.~R.},
        title = "{Quasi-periodic X-ray eruptions years after a nearby tidal disruption event}",
      journal = {\nat},
         year = 2024,
        month = oct,
       volume = {634},
       number = {8035},
        pages = {804-808},
          doi = {10.1038/s41586-024-08023-6},
archivePrefix = {arXiv},
       eprint = {2409.02181},
 primaryClass = {astro-ph.HE},
       adsurl = {https://ui.adsabs.harvard.edu/abs/2024Natur.634..804N}
}

@ARTICLE{Ohsuga+2005,
       author = {{Ohsuga}, Ken and {Mori}, Masao and {Nakamoto}, Taishi and {Mineshige}, Shin},
        title = "{Supercritical Accretion Flows around Black Holes: Two-dimensional, Radiation Pressure-dominated Disks with Photon Trapping}",
      journal = {\apj},
         year = 2005,
        month = jul,
       volume = {628},
       number = {1},
        pages = {368-381},
          doi = {10.1086/430728},
archivePrefix = {arXiv},
       eprint = {astro-ph/0504168},
 primaryClass = {astro-ph},
       adsurl = {https://ui.adsabs.harvard.edu/abs/2005ApJ...628..368O}
}

@ARTICLE{Olejak+2025,
       author = {{Olejak}, Aleksandra and {Stegmann}, Jakob and {de Mink}, Selma E. and {Valli}, Ruggero and {Sari}, Re'em and {Justham}, Stephen and {Ryu}, Taeho},
        title = "{Supermassive Black Holes Stripping a Subgiant Star Down to Its Helium Core: A New Type of Multimessenger Source for LISA}",
      journal = {\apjl},
         year = 2025,
        month = jul,
       volume = {987},
       number = {1},
          eid = {L11},
        pages = {L11},
          doi = {10.3847/2041-8213/ade432},
archivePrefix = {arXiv},
       eprint = {2503.21995},
 primaryClass = {astro-ph.HE},
       adsurl = {https://ui.adsabs.harvard.edu/abs/2025ApJ...987L..11O}
}

@ARTICLE{Pacucci&Narayan2024,
       author = {{Pacucci}, Fabio and {Narayan}, Ramesh},
        title = "{Mildly Super-Eddington Accretion onto Slowly Spinning Black Holes Explains the X-Ray Weakness of the Little Red Dots}",
      journal = {\apj},
         year = 2024,
        month = nov,
       volume = {976},
       number = {1},
          eid = {96},
        pages = {96},
          doi = {10.3847/1538-4357/ad84f7},
archivePrefix = {arXiv},
       eprint = {2407.15915},
 primaryClass = {astro-ph.HE},
       adsurl = {https://ui.adsabs.harvard.edu/abs/2024ApJ...976...96P}
}

@ARTICLE{Pan+2022,
       author = {{Pan}, Xin and {Li}, Shuang-Liang and {Cao}, Xinwu and {Miniutti}, Giovanni and {Gu}, Minfeng},
        title = "{A Disk Instability Model for the Quasi-periodic Eruptions of GSN 069}",
      journal = {\apjl},
         year = 2022,
        month = apr,
       volume = {928},
       number = {2},
          eid = {L18},
        pages = {L18},
          doi = {10.3847/2041-8213/ac5faf},
archivePrefix = {arXiv},
       eprint = {2203.12137},
 primaryClass = {astro-ph.GA},
       adsurl = {https://ui.adsabs.harvard.edu/abs/2022ApJ...928L..18P}
}

@ARTICLE{Pan+2023,
       author = {{Pan}, Xin and {Li}, Shuang-Liang and {Cao}, Xinwu},
        title = "{Application of the Disk Instability Model to All Quasiperiodic Eruptions}",
      journal = {\apj},
         year = 2023,
        month = jul,
       volume = {952},
       number = {1},
          eid = {32},
        pages = {32},
          doi = {10.3847/1538-4357/acd180},
archivePrefix = {arXiv},
       eprint = {2305.02071},
 primaryClass = {astro-ph.HE},
       adsurl = {https://ui.adsabs.harvard.edu/abs/2023ApJ...952...32P}
}

@ARTICLE{Pan+2025,
       author = {{Pan}, Xin and {Li}, Shuang-Liang and {Cao}, Xinwu and {Liu}, Bifang and {Yuan}, Weimin},
        title = "{Disk Instability Model for Quasi-periodic Eruptions: Investigating Period Dispersion and Peak Temperature}",
      journal = {\apj},
         year = 2025,
        month = aug,
       volume = {989},
       number = {2},
          eid = {196},
        pages = {196},
          doi = {10.3847/1538-4357/adf05d},
archivePrefix = {arXiv},
       eprint = {2507.11100},
 primaryClass = {astro-ph.HE},
       adsurl = {https://ui.adsabs.harvard.edu/abs/2025ApJ...989..196P}
}

@INPROCEEDINGS{Phinney1989,
       author = {{Phinney}, E.~S.},
        title = "{Manifestations of a Massive Black Hole in the Galactic Center}",
    booktitle = {The Center of the Galaxy},
         year = 1989,
       editor = {{Morris}, Mark},
       series = {IAU Symposium},
       volume = {136},
        month = jan,
        pages = {543},
       adsurl = {https://ui.adsabs.harvard.edu/abs/1989IAUS..136..543P}
}

@ARTICLE{Piran+2015,
       author = {{Piran}, Tsvi and {Svirski}, Gilad and {Krolik}, Julian and {Cheng}, Roseanne M. and {Shiokawa}, Hotaka},
        title = "{‧Disk Formation Versus Disk Accretion{\textemdash}What Powers Tidal Disruption Events?}",
      journal = {\apj},
         year = 2015,
        month = jun,
       volume = {806},
       number = {2},
          eid = {164},
        pages = {164},
          doi = {10.1088/0004-637X/806/2/164},
archivePrefix = {arXiv},
       eprint = {1502.05792},
 primaryClass = {astro-ph.HE},
       adsurl = {https://ui.adsabs.harvard.edu/abs/2015ApJ...806..164P}
}

@ARTICLE{Piro&Mockler2025,
       author = {{Piro}, Anthony L. and {Mockler}, Brenna},
        title = "{Late-time Evolution and Instabilities of Tidal Disruption Disks}",
      journal = {\apj},
         year = 2025,
        month = may,
       volume = {985},
       number = {1},
          eid = {77},
        pages = {77},
          doi = {10.3847/1538-4357/adc729},
archivePrefix = {arXiv},
       eprint = {2412.01922},
 primaryClass = {astro-ph.HE},
       adsurl = {https://ui.adsabs.harvard.edu/abs/2025ApJ...985...77P}
}

@ARTICLE{Price+2024,
       author = {{Price}, Daniel J. and {Liptai}, David and {Mandel}, Ilya and {Shepherd}, Joanna and {Lodato}, Giuseppe and {Levin}, Yuri},
        title = "{Eddington Envelopes: The Fate of Stars on Parabolic Orbits Tidally Disrupted by Supermassive Black Holes}",
      journal = {\apjl},
         year = 2024,
        month = aug,
       volume = {971},
       number = {2},
          eid = {L46},
        pages = {L46},
          doi = {10.3847/2041-8213/ad6862},
archivePrefix = {arXiv},
       eprint = {2404.09381},
 primaryClass = {astro-ph.HE},
       adsurl = {https://ui.adsabs.harvard.edu/abs/2024ApJ...971L..46P}
}

@PHDTHESIS{Pringle1974,
       author = {{Pringle}, James Edward},
        title = "{Stellar radial migration and thick disks in spiral galaxies}",
       school = {Rutgers University, New Jersey},
         year = 1974,
        month = sep,
       adsurl = {https://ui.adsabs.harvard.edu/abs/1974PhDT.......131P}
}

@ARTICLE{Pringle1976,
       author = {{Pringle}, J.~E.},
        title = "{Thermal instabilities in accretion discs.}",
      journal = {\mnras},
         year = 1976,
        month = oct,
       volume = {177},
        pages = {65-71},
          doi = {10.1093/mnras/177.1.65},
       adsurl = {https://ui.adsabs.harvard.edu/abs/1976MNRAS.177...65P}
}

@ARTICLE{Pringle1991,
       author = {{Pringle}, J.~E.},
        title = "{The properties of external accretion discs.}",
      journal = {\mnras},
         year = 1991,
        month = feb,
       volume = {248},
        pages = {754},
          doi = {10.1093/mnras/248.4.754},
       adsurl = {https://ui.adsabs.harvard.edu/abs/1991MNRAS.248..754P}
}

@ARTICLE{Quintin+2023,
       author = {{Quintin}, E. and {Webb}, N.~A. and {Guillot}, S. and {Miniutti}, G. and {Kammoun}, E.~S. and {Giustini}, M. and {Arcodia}, R. and {Soucail}, G. and {Clerc}, N. and {Amato}, R. and {Markwardt}, C.~B.},
        title = "{Tormund's return: Hints of quasi-periodic eruption features from a recent optical tidal disruption event}",
      journal = {\aap},
         year = 2023,
        month = jul,
       volume = {675},
          eid = {A152},
        pages = {A152},
          doi = {10.1051/0004-6361/202346440},
archivePrefix = {arXiv},
       eprint = {2306.00438},
 primaryClass = {astro-ph.HE},
       adsurl = {https://ui.adsabs.harvard.edu/abs/2023A&A...675A.152Q}
}

@ARTICLE{Raj&Nixon2021,
       author = {{Raj}, A. and {Nixon}, C.~J.},
        title = "{Disk Tearing: Implications for Black Hole Accretion and AGN Variability}",
      journal = {\apj},
         year = 2021,
        month = mar,
       volume = {909},
       number = {1},
          eid = {82},
        pages = {82},
          doi = {10.3847/1538-4357/abdc25},
archivePrefix = {arXiv},
       eprint = {2101.05825},
 primaryClass = {astro-ph.HE},
       adsurl = {https://ui.adsabs.harvard.edu/abs/2021ApJ...909...82R}
}

@ARTICLE{Ramillard+2022,
       author = {{Remillard}, Ronald A. and {Loewenstein}, Michael and {Steiner}, James F. and {Prigozhin}, Gregory Y. and {LaMarr}, Beverly and {Enoto}, Teruaki and {Gendreau}, Keith C. and {Arzoumanian}, Zaven and {Markwardt}, Craig and {Basak}, Arkadip and {Stevens}, Abigail L. and {Ray}, Paul S. and {Altamirano}, Diego and {Buisson}, Douglas J.~K.},
        title = "{An Empirical Background Model for the NICER X-Ray Timing Instrument}",
      journal = {\aj},
         year = 2022,
        month = mar,
       volume = {163},
       number = {3},
          eid = {130},
        pages = {130},
          doi = {10.3847/1538-3881/ac4ae6},
archivePrefix = {arXiv},
       eprint = {2105.09901},
 primaryClass = {astro-ph.IM},
       adsurl = {https://ui.adsabs.harvard.edu/abs/2022AJ....163..130R}
}

@ARTICLE{Rees1988,
       author = {{Rees}, Martin J.},
        title = "{Tidal disruption of stars by black holes of {}10$^{6}$-{}10$^{8}$ solar masses in nearby galaxies}",
      journal = {\nat},
         year = 1988,
        month = jun,
       volume = {333},
       number = {6173},
        pages = {523-528},
          doi = {10.1038/333523a0},
       adsurl = {https://ui.adsabs.harvard.edu/abs/1988Natur.333..523R}
}

@BOOK{Rybicki&Lightman1986,
       author = {{Rybicki}, George B. and {Lightman}, Alan P.},
        title = "{Radiative Processes in Astrophysics}",
         year = 1986,
       adsurl = {https://ui.adsabs.harvard.edu/abs/1986rpa..book.....R}
}

@ARTICLE{Ryu+2023b,
       author = {{Ryu}, Taeho and {Krolik}, Julian and {Piran}, Tsvi and {Noble}, Scott C. and {Avara}, Mark},
        title = "{Shocks Power Tidal Disruption Events}",
      journal = {\apj},
         year = 2023,
        month = nov,
       volume = {957},
       number = {1},
          eid = {12},
        pages = {12},
          doi = {10.3847/1538-4357/acf5de},
archivePrefix = {arXiv},
       eprint = {2305.05333},
 primaryClass = {astro-ph.HE},
       adsurl = {https://ui.adsabs.harvard.edu/abs/2023ApJ...957...12R}
}

@ARTICLE{Sadowski+2016,
       author = {{S{\k{a}}dowski}, Aleksander and {Narayan}, Ramesh},
        title = "{Three-dimensional simulations of supercritical black hole accretion discs - luminosities, photon trapping and variability}",
      journal = {\mnras},
         year = 2016,
        month = mar,
       volume = {456},
       number = {4},
        pages = {3929-3947},
          doi = {10.1093/mnras/stv2941},
archivePrefix = {arXiv},
       eprint = {1509.03168},
 primaryClass = {astro-ph.HE},
       adsurl = {https://ui.adsabs.harvard.edu/abs/2016MNRAS.456.3929S}
}

@ARTICLE{Sazonov+2021,
       author = {{Sazonov}, S. and {Gilfanov}, M. and {Medvedev}, P. and {Yao}, Y. and {Khorunzhev}, G. and {Semena}, A. and {Sunyaev}, R. and {Burenin}, R. and {Lyapin}, A. and {Meshcheryakov}, A. and {Uskov}, G. and {Zaznobin}, I. and {Postnov}, K.~A. and {Dodin}, A.~V. and {Belinski}, A.~A. and {Cherepashchuk}, A.~M. and {Eselevich}, M. and {Dodonov}, S.~N. and {Grokhovskaya}, A.~A. and {Kotov}, S.~S. and {Bikmaev}, I.~F. and {Zhuchkov}, R. Ya and {Gumerov}, R.~I. and {van Velzen}, S. and {Kulkarni}, S.},
        title = "{First tidal disruption events discovered by SRG/eROSITA: X-ray/optical properties and X-ray luminosity function at z < 0.6}",
      journal = {\mnras},
         year = 2021,
        month = dec,
       volume = {508},
       number = {3},
        pages = {3820-3847},
          doi = {10.1093/mnras/stab2843},
archivePrefix = {arXiv},
       eprint = {2108.02449},
 primaryClass = {astro-ph.HE},
       adsurl = {https://ui.adsabs.harvard.edu/abs/2021MNRAS.508.3820S}
}

@ARTICLE{Shakura&Sunyaev1973,
       author = {{Shakura}, N.~I. and {Sunyaev}, R.~A.},
        title = "{Black holes in binary systems. Observational appearance.}",
      journal = {\aap},
         year = 1973,
        month = jan,
       volume = {24},
        pages = {337-355},
       adsurl = {https://ui.adsabs.harvard.edu/abs/1973A&A....24..337S}
}

@ARTICLE{Shakura&Sunyaev1976,
       author = {{Shakura}, N.~I. and {Sunyaev}, R.~A.},
        title = "{A theory of the instability of disk accretion on to black holes and the variability of binary X-ray sources, galactic nuclei and quasars.}",
      journal = {\mnras},
         year = 1976,
        month = jun,
       volume = {175},
        pages = {613-632},
          doi = {10.1093/mnras/175.3.613},
       adsurl = {https://ui.adsabs.harvard.edu/abs/1976MNRAS.175..613S}
}

@ARTICLE{Shen&Matzner2014,
       author = {{Shen}, Rong-Feng and {Matzner}, Christopher D.},
        title = "{Evolution of Accretion Disks in Tidal Disruption Events}",
      journal = {\apj},
         year = 2014,
        month = apr,
       volume = {784},
       number = {2},
          eid = {87},
        pages = {87},
          doi = {10.1088/0004-637X/784/2/87},
archivePrefix = {arXiv},
       eprint = {1305.5570},
 primaryClass = {astro-ph.HE},
       adsurl = {https://ui.adsabs.harvard.edu/abs/2014ApJ...784...87S}
}

@ARTICLE{Shibazaki&Hoshi1975,
       author = {{Shibazaki}, N. and {H{\={o}}shi}, R.},
        title = "{Structure and Stability of Accretion-Disk around a Black-Hole}",
      journal = {Progress of Theoretical Physics},
         year = 1975,
        month = sep,
       volume = {54},
       number = {3},
        pages = {706-718},
          doi = {10.1143/PTP.54.706},
       adsurl = {https://ui.adsabs.harvard.edu/abs/1975PThPh..54..706S}
}

@ARTICLE{Shiokawa+2015,
       author = {{Shiokawa}, Hotaka and {Krolik}, Julian H. and {Cheng}, Roseanne M. and {Piran}, Tsvi and {Noble}, Scott C.},
        title = "{General Relativistic Hydrodynamic Simulation of Accretion Flow from a Stellar Tidal Disruption}",
      journal = {\apj},
         year = 2015,
        month = may,
       volume = {804},
       number = {2},
          eid = {85},
        pages = {85},
          doi = {10.1088/0004-637X/804/2/85},
archivePrefix = {arXiv},
       eprint = {1501.04365},
 primaryClass = {astro-ph.HE},
       adsurl = {https://ui.adsabs.harvard.edu/abs/2015ApJ...804...85S}
}

@ARTICLE{Sniegowska+2023,
       author = {{{\'S}niegowska}, Marzena and {Grz{\c{e}}dzielski}, Miko{\l}aj and {Czerny}, Bo{\.z}ena and {Janiuk}, Agnieszka},
        title = "{Modified models of radiation pressure instability applied to 10, {}10$^{5}$, and {}10$^{7}$ M$_{{\ensuremath{\odot}}}$ accreting black holes}",
      journal = {\aap},
         year = 2023,
        month = apr,
       volume = {672},
          eid = {A19},
        pages = {A19},
          doi = {10.1051/0004-6361/202243828},
archivePrefix = {arXiv},
       eprint = {2204.10067},
 primaryClass = {astro-ph.HE},
       adsurl = {https://ui.adsabs.harvard.edu/abs/2023A&A...672A..19S}
}

@ARTICLE{Steinberg&Stone2024,
       author = {{Steinberg}, Elad and {Stone}, Nicholas C.},
        title = "{Stream-disk shocks as the origins of peak light in tidal disruption events}",
      journal = {\nat},
         year = 2024,
        month = jan,
       volume = {625},
       number = {7995},
        pages = {463-467},
          doi = {10.1038/s41586-023-06875-y},
archivePrefix = {arXiv},
       eprint = {2206.10641},
 primaryClass = {astro-ph.HE},
       adsurl = {https://ui.adsabs.harvard.edu/abs/2024Natur.625..463S}
}

@ARTICLE{Strubbe&Quataert2009,
       author = {{Strubbe}, Linda E. and {Quataert}, Eliot},
        title = "{Optical flares from the tidal disruption of stars by massive black holes}",
      journal = {\mnras},
         year = 2009,
        month = dec,
       volume = {400},
       number = {4},
        pages = {2070-2084},
          doi = {10.1111/j.1365-2966.2009.15599.x},
archivePrefix = {arXiv},
       eprint = {0905.3735},
 primaryClass = {astro-ph.CO},
       adsurl = {https://ui.adsabs.harvard.edu/abs/2009MNRAS.400.2070S}
}

@ARTICLE{Sukova+2021,
       author = {{Sukov{\'a}}, Petra and {Zaja{\v{c}}ek}, Michal and {Witzany}, Vojt{\v{e}}ch and {Karas}, Vladim{\'\i}r},
        title = "{Stellar Transits across a Magnetized Accretion Torus as a Mechanism for Plasmoid Ejection}",
      journal = {\apj},
         year = 2021,
        month = aug,
       volume = {917},
       number = {1},
          eid = {43},
        pages = {43},
          doi = {10.3847/1538-4357/ac05c6},
archivePrefix = {arXiv},
       eprint = {2102.08135},
 primaryClass = {astro-ph.HE},
       adsurl = {https://ui.adsabs.harvard.edu/abs/2021ApJ...917...43S}
}

@ARTICLE{Suzuguchi+2025,
       author = {{Suzuguchi}, Tomoya and {Omiya}, Hidetoshi and {Takeda}, Hiroki},
        title = "{Possibility of Multi-Messenger Observations of Quasi-Periodic Eruptions with X-rays and Gravitational Waves}",
      journal = {arXiv e-prints},
         year = 2025,
        month = may,
          eid = {arXiv:2505.10488},
        pages = {arXiv:2505.10488},
          doi = {10.48550/arXiv.2505.10488},
archivePrefix = {arXiv},
       eprint = {2505.10488},
 primaryClass = {astro-ph.HE},
       adsurl = {https://ui.adsabs.harvard.edu/abs/2025arXiv250510488S}
}

@ARTICLE{Tagawa&Haiman2023,
       author = {{Tagawa}, Hiromichi and {Haiman}, Zolt{\'a}n},
        title = "{Flares from stars crossing active galactic nucleus discs on low-inclination orbits}",
      journal = {\mnras},
         year = 2023,
        month = nov,
       volume = {526},
       number = {1},
        pages = {69-79},
          doi = {10.1093/mnras/stad2616},
archivePrefix = {arXiv},
       eprint = {2304.03670},
 primaryClass = {astro-ph.HE},
       adsurl = {https://ui.adsabs.harvard.edu/abs/2023MNRAS.526...69T}
}

@ARTICLE{Tamilan+2025,
       author = {{Tamilan}, Mageshwaran and {Hayasaki}, Kimitake and {Suzuki}, Takeru K.},
        title = "{Self-Similar Solutions for Geometrically Thin Accretion Disks with Magnetically Driven Winds: Application to Tidal Disruption Events}",
      journal = {arXiv e-prints},
         year = 2025,
        month = feb,
          eid = {arXiv:2502.12549},
        pages = {arXiv:2502.12549},
          doi = {10.48550/arXiv.2502.12549},
archivePrefix = {arXiv},
       eprint = {2502.12549},
 primaryClass = {astro-ph.HE},
       adsurl = {https://ui.adsabs.harvard.edu/abs/2025arXiv250212549T}
}

@ARTICLE{Thomsen+2022,
       author = {{Thomsen}, Lars L. and {Kwan}, Tom M. and {Dai}, Lixin and {Wu}, Samantha C. and {Roth}, Nathaniel and {Ramirez-Ruiz}, Enrico},
        title = "{Dynamical Unification of Tidal Disruption Events}",
      journal = {\apjl},
         year = 2022,
        month = oct,
       volume = {937},
       number = {2},
          eid = {L28},
        pages = {L28},
          doi = {10.3847/2041-8213/ac911f},
archivePrefix = {arXiv},
       eprint = {2206.02804},
 primaryClass = {astro-ph.HE},
       adsurl = {https://ui.adsabs.harvard.edu/abs/2022ApJ...937L..28T}
}

@ARTICLE{Tsz-Lok_Lam+2025,
       author = {{Tsz-Lok Lam}, Alan and {Shibata}, Masaru and {Kawaguchi}, Kyohei and {Pelle}, Joaquin},
        title = "{Black hole-accretion disk collision in general relativity: Axisymmetric simulations}",
      journal = {arXiv e-prints},
         year = 2025,
        month = apr,
          eid = {arXiv:2504.17016},
        pages = {arXiv:2504.17016},
          doi = {10.48550/arXiv.2504.17016},
archivePrefix = {arXiv},
       eprint = {2504.17016},
 primaryClass = {astro-ph.HE},
       adsurl = {https://ui.adsabs.harvard.edu/abs/2025arXiv250417016T}
}

@ARTICLE{van_Velzen+2019,
       author = {{van Velzen}, Sjoert and {Stone}, Nicholas C. and {Metzger}, Brian D. and {Gezari}, Suvi and {Brown}, Thomas M. and {Fruchter}, Andrew S.},
        title = "{Late-time UV Observations of Tidal Disruption Flares Reveal Unobscured, Compact Accretion Disks}",
      journal = {\apj},
         year = 2019,
        month = jun,
       volume = {878},
       number = {2},
          eid = {82},
        pages = {82},
          doi = {10.3847/1538-4357/ab1844},
archivePrefix = {arXiv},
       eprint = {1809.00003},
 primaryClass = {astro-ph.HE},
       adsurl = {https://ui.adsabs.harvard.edu/abs/2019ApJ...878...82V}
}

@ARTICLE{Vurm+2025,
       author = {{Vurm}, Indrek and {Linial}, Itai and {Metzger}, Brian D.},
        title = "{Radiation Transport Simulations of Quasiperiodic Eruptions from Star{\textendash}Disk Collisions}",
      journal = {\apj},
         year = 2025,
        month = apr,
       volume = {983},
       number = {1},
          eid = {40},
        pages = {40},
          doi = {10.3847/1538-4357/adb74d},
archivePrefix = {arXiv},
       eprint = {2410.05166},
 primaryClass = {astro-ph.HE},
       adsurl = {https://ui.adsabs.harvard.edu/abs/2025ApJ...983...40V}
}

@ARTICLE{Wang&Zhou1999,
       author = {{Wang}, Jian-Min and {Zhou}, You-Yuan},
        title = "{Self-similar Solution of Optically Thick Advection-dominated Flows}",
      journal = {\apj},
         year = 1999,
        month = may,
       volume = {516},
       number = {1},
        pages = {420-424},
          doi = {10.1086/307080},
       adsurl = {https://ui.adsabs.harvard.edu/abs/1999ApJ...516..420W}
}

@ARTICLE{Watarai&Fukue1999,
       author = {{Watarai}, Ken-ya and {Fukue}, Jun},
        title = "{Radiative Disk Winds from a Self-Similar Slim Disk}",
      journal = {\pasj},
         year = 1999,
        month = oct,
       volume = {51},
        pages = {725},
          doi = {10.1093/pasj/51.5.725},
       adsurl = {https://ui.adsabs.harvard.edu/abs/1999PASJ...51..725W}
}

@ARTICLE{Watarai2006,
       author = {{Watarai}, Ken-ya},
        title = "{New Analytical Formulae for Supercritical Accretion Flows}",
      journal = {\apj},
         year = 2006,
        month = sep,
       volume = {648},
       number = {1},
        pages = {523-533},
          doi = {10.1086/505854},
archivePrefix = {arXiv},
       eprint = {astro-ph/0605248},
 primaryClass = {astro-ph},
       adsurl = {https://ui.adsabs.harvard.edu/abs/2006ApJ...648..523W}
}

@ARTICLE{Wevers+2022,
       author = {{Wevers}, T. and {Pasham}, D.~R. and {Jalan}, P. and {Rakshit}, S. and {Arcodia}, R.},
        title = "{Host galaxy properties of quasi-periodically erupting X-ray sources}",
      journal = {\aap},
         year = 2022,
        month = mar,
       volume = {659},
          eid = {L2},
        pages = {L2},
          doi = {10.1051/0004-6361/202243143},
archivePrefix = {arXiv},
       eprint = {2201.11751},
 primaryClass = {astro-ph.HE},
       adsurl = {https://ui.adsabs.harvard.edu/abs/2022A&A...659L...2W}
}

@ARTICLE{Wevers+2024,
       author = {{Wevers}, T. and {French}, K.~D. and {Zabludoff}, A.~I. and {Fischer}, T.~C. and {Rowlands}, K. and {Guolo}, M. and {Dalla Barba}, B. and {Arcodia}, R. and {Berton}, M. and {Bian}, F. and {Linial}, I. and {Miniutti}, G. and {Pasham}, D.~R.},
        title = "{X-Ray Quasi-periodic Eruptions and Tidal Disruption Events Prefer Similar Host Galaxies}",
      journal = {\apjl},
         year = 2024,
        month = jul,
       volume = {970},
       number = {1},
          eid = {L23},
        pages = {L23},
          doi = {10.3847/2041-8213/ad5f1b},
archivePrefix = {arXiv},
       eprint = {2406.02678},
 primaryClass = {astro-ph.HE},
       adsurl = {https://ui.adsabs.harvard.edu/abs/2024ApJ...970L..23W}
}

@ARTICLE{Xian+2021,
       author = {{Xian}, Jingtao and {Zhang}, Fupeng and {Dou}, Liming and {He}, Jiasheng and {Shu}, Xinwen},
        title = "{X-Ray Quasi-periodic Eruptions Driven by Star-Disk Collisions: Application to GSN069 and Probing the Spin of Massive Black Holes}",
      journal = {\apjl},
         year = 2021,
        month = nov,
       volume = {921},
       number = {2},
          eid = {L32},
        pages = {L32},
          doi = {10.3847/2041-8213/ac31aa},
archivePrefix = {arXiv},
       eprint = {2110.10855},
 primaryClass = {astro-ph.HE},
       adsurl = {https://ui.adsabs.harvard.edu/abs/2021ApJ...921L..32X}
}

@ARTICLE{Yao+2025,
       author = {{Yao}, Philippe Z. and {Quataert}, Eliot and {Jiang}, Yan-Fei and {Lu}, Wenbin and {White}, Christopher J.},
        title = "{Star‑Disk Collisions: Implications for Quasi-periodic Eruptions and Other Transients near Supermassive Black Holes}",
      journal = {\apj},
         year = 2025,
        month = jan,
       volume = {978},
       number = {1},
          eid = {91},
        pages = {91},
          doi = {10.3847/1538-4357/ad8911},
archivePrefix = {arXiv},
       eprint = {2407.14578},
 primaryClass = {astro-ph.HE},
       adsurl = {https://ui.adsabs.harvard.edu/abs/2025ApJ...978...91Y}
}

@ARTICLE{Yao&Quataert2025,
       author = {{Yao}, Philippe Z. and {Quataert}, Eliot},
        title = "{Mass Transfer in Tidally Heated Stars Orbiting Massive Black Holes and Implications for Repeating Nuclear Transients}",
      journal = {arXiv e-prints},
         year = 2025,
        month = may,
          eid = {arXiv:2505.10611},
        pages = {arXiv:2505.10611},
          doi = {10.48550/arXiv.2505.10611},
archivePrefix = {arXiv},
       eprint = {2505.10611},
 primaryClass = {astro-ph.HE},
       adsurl = {https://ui.adsabs.harvard.edu/abs/2025arXiv250510611Y}
}

@ARTICLE{Zhou+2024a,
       author = {{Zhou}, Cong and {Huang}, Lei and {Guo}, Kangrou and {Li}, Ya-Ping and {Pan}, Zhen},
        title = "{Probing orbits of stellar mass objects deep in galactic nuclei with quasiperiodic eruptions}",
      journal = {\prd},
         year = 2024,
        month = may,
       volume = {109},
       number = {10},
          eid = {103031},
        pages = {103031},
          doi = {10.1103/PhysRevD.109.103031},
archivePrefix = {arXiv},
       eprint = {2401.11190},
 primaryClass = {astro-ph.HE},
       adsurl = {https://ui.adsabs.harvard.edu/abs/2024PhRvD.109j3031Z}
}

@ARTICLE{Zhou+2024b,
       author = {{Zhou}, Cong and {Zhong}, Binyu and {Zeng}, Yuhe and {Huang}, Lei and {Pan}, Zhen},
        title = "{Probing orbits of stellar mass objects deep in galactic nuclei with quasiperiodic eruptions. II. Population analysis}",
      journal = {\prd},
         year = 2024,
        month = oct,
       volume = {110},
       number = {8},
          eid = {083019},
        pages = {083019},
          doi = {10.1103/PhysRevD.110.083019},
archivePrefix = {arXiv},
       eprint = {2405.06429},
 primaryClass = {astro-ph.HE},
       adsurl = {https://ui.adsabs.harvard.edu/abs/2024PhRvD.110h3019Z}
}

@ARTICLE{Zhou+2025,
       author = {{Zhou}, Cong and {Zeng}, Yuhe and {Pan}, Zhen},
        title = "{Secular Evolution of Quasiperiodic Eruptions}",
      journal = {\apj},
         year = 2025,
        month = jun,
       volume = {985},
       number = {2},
          eid = {242},
        pages = {242},
          doi = {10.3847/1538-4357/adcee2},
archivePrefix = {arXiv},
       eprint = {2411.18046},
 primaryClass = {astro-ph.HE},
       adsurl = {https://ui.adsabs.harvard.edu/abs/2025ApJ...985..242Z}
}

% Alternatively you could enter them by hand, like this:
% This method is tedious and prone to error if you have lots of references
%\begin{thebibliography}{99}
%\bibitem[\protect\citeauthoryear{Author}{2012}]{Author2012}
%Author A.~N., 2013, Journal of Improbable Astronomy, 1, 1
%\bibitem[\protect\citeauthoryear{Others}{2013}]{Others2013}
%Others S., 2012, Journal of Interesting Stuff, 17, 198
%\end{thebibliography}

%%%%%%%%%%%%%%%%%%%%%%%%%%%%%%%%%%%%%%%%%%%%%%%%%%

%%%%%%%%%%%%%%%%% APPENDICES %%%%%%%%%%%%%%%%%%%%%

\appendix

\section{Self-Similar Solution} \label{appsec:disk}
We summarize the self-similar solution by \citet{Cannizzo&Gehrels2009}. The time evolution of the surface density is determined by the diffusion equation:
\begin{align}
    \label{appeq:Sigma_evolution}
    \frac{\partial\Sigma}{\partial t} &= \frac{3}{R}\frac{\partial}{\partial R}
    \left[R^{1/2}\frac{\partial}{\partial R}(\nu_{\rm vis}\Sigma R^{1/2})\right]\ ,
\end{align}
which follows from the mass and angular momentum conservations:
\begin{align}
    &\frac{\partial \Sigma}{\partial t}
    = \frac{1}{2 \pi R} \frac{\partial \dot{M}}{\partial R}\ ,\\
    &
    \label{appeq:ang cons}
    -\frac{\dot{M}}{2 \pi} \frac{{\rm d}}{{\rm d} R}
    \left(R^{2} \Omega\right)
    = \frac{\partial}{\partial R}
    \left(\nu_{\rm vis} \Sigma R^{3}\frac{{\rm d}\Omega}{{\rm d}R}\right)\ ,
\end{align}
and the Keplerian rotation ($\Omega=\Omega_{\rm K}=\sqrt{GM_{\bullet}/R^{3}}$) is assumed \citep[e.g.,][]{Kato+2008}. Here the viscosity is given by the so-called $\alpha$ prescription by \cite{Shakura&Sunyaev1973}
\begin{align}
    \label{appeq:nu}
    \nu_{\rm vis} = \frac{2 \alpha P}{3 \Omega_{\rm K} \rho}
    \simeq{\cal C} R^{1/2}
    {\,\,\,\rm and \,\,\,}
    {\cal C} \equiv \frac{2}{3} \alpha (G M_{\bullet})^{1/2}\ ,
\end{align}
where $\rho$ and $c_{\rm s} \equiv (P/\rho)^{1/2}$ is the (midplane) density and the sound velocity, respectively. In the second equality, the vertical hydrostatic equilibrium ($c_{\rm s} \simeq H \Omega_{\rm K}$), and the geometrically thick disk
\begin{align}
    \label{appeq:scale_height_ss}
    H &\simeq R\ ,
\end{align}
are assumed. For the viscosity of Eq.~\eqref{appeq:nu}, Eq.~\eqref{appeq:Sigma_evolution} has a self-similar solution conserving the angular momentum \citep{Lyubarskij&Shakura1987,Pringle1991}:
\begin{align}
    \label{appeq:self_similar_sigma}
    \Sigma(R,t) &= \Sigma_{0} 
    \left(\frac{R}{R_{0}}\right)^{-1/2}
    \left(\frac{t}{t_{0}}\right)^{-4/3} \nonumber \\
    &\quad \times {\rm exp}
    \left[-\frac{1}{9}\left(\frac{R}{R_{0}}\right)^{3/2}\left(\frac{t}{t_{0}}\right)^{-1}\right],
\end{align}
where $\Sigma_{0}$, $R_{0}$, and $t_{0}$ are arbitrary normalization constant. The normalization of radius and time should satisfy the closure relation
\begin{align}
    \label{appeq:closure_relation}
    t_{0} &= \frac{4R_{0}^{3/2}}{3{\cal C}}\ .
\end{align}

The disk mass is given by
\begin{align}
    M_{\rm disk} = 
    \int_{0}^{\infty} 2 \pi R \Sigma {\rm d}R
    = 12 \pi R_{0}^2 \Sigma_{0} 
    \left(\frac{t}{t_{0}}\right)^{-1/3}\ ,
\end{align}
and its time derivative gives the (global) accretion rate onto the center
\begin{align}
    \frac{{\rm d}M}{{\rm d}t}
    = \frac{4 \pi R_{0}^{2} \Sigma_{0}}{t_{0}}
    \left(\frac{t}{t_{0}}\right)^{-4/3}\ ,
\end{align}
where we define ${\rm d}M/{\rm d}t \equiv |{\rm d}M_{\rm disk}/{\rm d}t|$. On the other hand, the local accretion rate is given by 
Eq.~\eqref{appeq:ang cons}
\begin{align}
    \dot{M}(R,t) &= \frac{{\rm d}M}{{\rm d}t} 
    \left[1-\frac{1}{3}\left(\frac{R}{R_{0}}\right)^{\frac{3}{2}}
    \left(\frac{t}{t_{0}}\right)^{-1}\right] \nonumber \\ 
    &\quad \times {\rm exp}
    \left[-\frac{1}{9}\left(\frac{R}{R_{0}}\right)^{\frac{3}{2}}
    \left(\frac{t}{t_{0}}\right)^{-1}\right]\ .
\end{align}

\section{Parameter Dependencies for Self-Similar Model} \label{appsec:parameter_ss}
\begin{figure*}
    \centering
    \includegraphics[keepaspectratio, scale=0.3,bb=0 0 1728 1296]{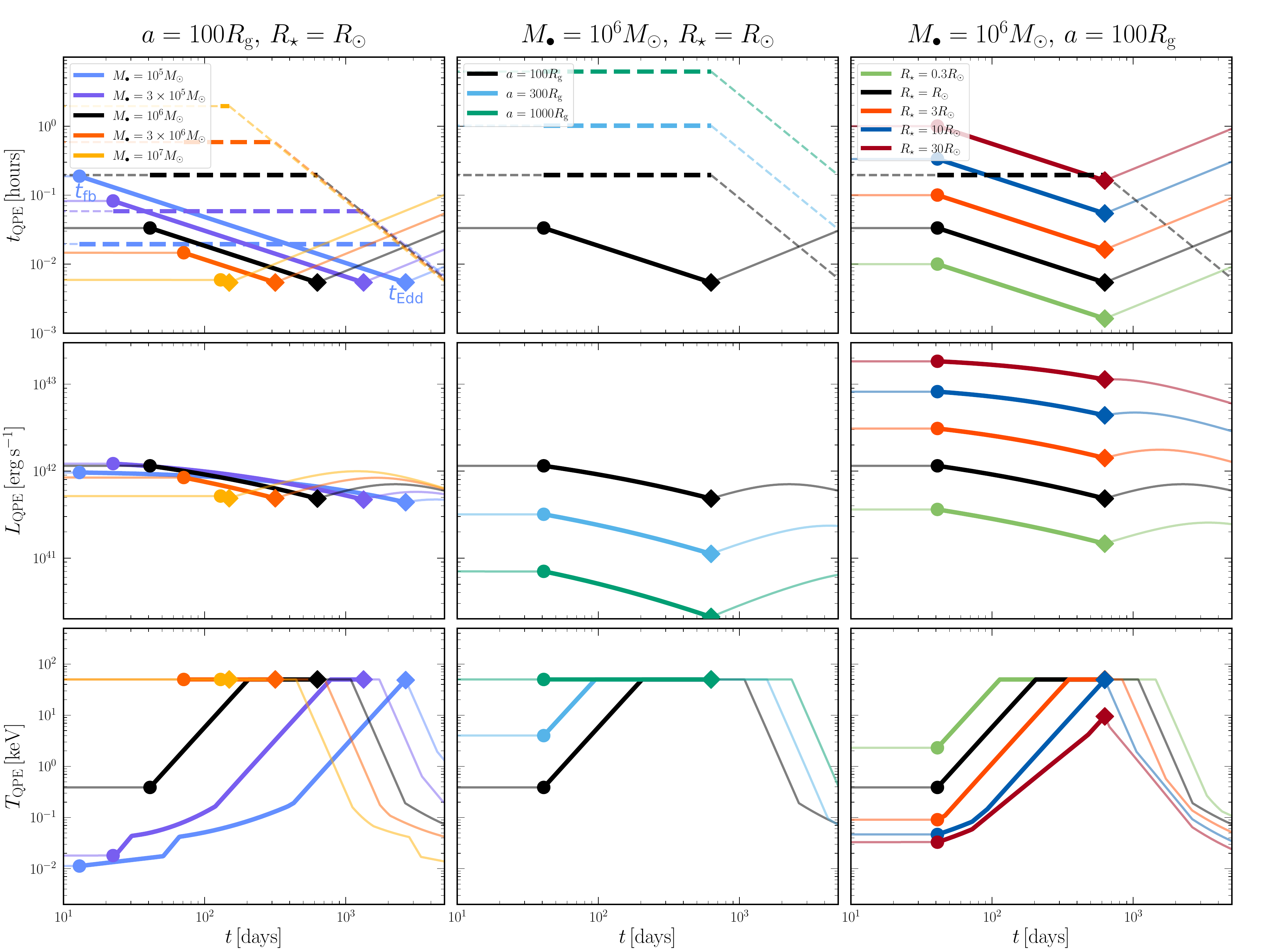}
    \caption{Same as Fig.~\ref{fig:parameter}, but for the SS model. On each curves, the moment of $t = t_{\rm fb}$ and $t = t_{\rm Edd}$ are shown by circles and diamonds.}
    \label{fig:parameter_ss}
\end{figure*}

Figure~\ref{fig:parameter_ss} shows the parameter dependencies of the QPE observables for the SS model. While the overall behavior is similar to that of the SQ model, the temperature is higher than the SQ model because of the lower surface density.

%%%%%%%%%%%%%%%%%%%%%%%%%%%%%%%%%%%%%%%%%%%%%%%%%%

% Don't change these lines
\bsp	% typesetting comment
\label{lastpage}
\end{document}